\documentclass[twocolumn,showkeys]{revtex4-1}

\pdfoutput=1
\usepackage[utf8]{inputenc}
\usepackage{bbm}
\usepackage{appendix}
\usepackage[frozencache,cachedir=.]{minted}

\input{default-pyg-prefix.pygstyle}
\input{default.pygstyle}

\usepackage[usenames,dvipsnames]{xcolor}
\usepackage{color}
\usepackage{colortbl}

\definecolor{qctrl_primary}{HTML}{680Ce9}
\definecolor{qctrl_secondary}{HTML}{BF04DC}
\definecolor{qctrl_noise}{HTML}{7B7479}
\definecolor{qctrl_axis_labels}{HTML}{514B4F}
\definecolor{qctrl_borders}{HTML}{CFCBCE}
\definecolor{qctrl_blue}{HTML}{4177D8}
\definecolor{qctrl_aqua}{HTML}{32A4A8}
\definecolor{qctrl_green}{HTML}{32A857}
\definecolor{qctrl_lime_green}{HTML}{A2A933}
\definecolor{qctrl_orange}{HTML}{D6742F}
\definecolor{qctrl_red}{HTML}{D84144}
\definecolor{qctrl_fuchsia}{HTML}{D84190}

\usepackage{blindtext}

\usepackage{amssymb,amsmath,amsthm}
\usepackage{mathtools} 

\usepackage{mathrsfs}
\usepackage{dsfont} 
\usepackage[nice]{nicefrac}

\usepackage{cancel}
\usepackage{bm}

\usepackage{algorithm}
\usepackage{algpseudocode}

\theoremstyle{definition}

\theoremstyle{remark}

\DeclareMathOperator{\sinc}{sinc}

\usepackage{physics}
\usepackage{siunitx} 
\usepackage{qcircuit}
\usepackage{braket}

\usepackage{graphicx}
\usepackage{float}
\usepackage{rotating}

\usepackage{dcolumn}
\usepackage{multirow}
\usepackage{makecell}
\usepackage{bigdelim}
\usepackage{blkarray}
\usepackage{array}

\usepackage{hyperref}
\hypersetup{
final=true,
plainpages=false,
pdfstartview=FitV,
pdftoolbar=true,
pdfmenubar=true,
bookmarksopen=true,
bookmarksnumbered=true,
breaklinks=true,
linktocpage,
colorlinks=true,
linkcolor=blue,
filecolor=blue,      
urlcolor=cyan,
citecolor=blue,
anchorcolor=green
}

\usepackage[nameinlink]{cleveref}
\crefname{equation}{Eq.}{Eqs.}
\crefname{align}{Eq.}{Eqs.}
\crefname{figure}{Fig.}{Figs.}
\crefname{table}{Table}{Tables}
\crefname{tabular}{Table}{Tables}
\crefname{section}{Sec.}{Secs.}
\crefname{appendix}{App.}{Apps.}

\crefname{appsec}{App.}{Apps.}
\crefname{appchapter}{App.}{Apps.}

\crefname{algorithm}{Algo.}{Algos.}
\creflabelformat{equation}{#2#1#3}

\usepackage{natbib}
\usepackage{diagbox}
\usepackage{array}
\usepackage{booktabs}

\begin{document}

\title{Error-robust quantum logic optimization using a cloud quantum computer interface}
\author{Andre R. R. Carvalho}
\author{Harrison Ball}
\author{Michael J. Biercuk}
\author{Michael R. Hush}
\author{Felix Thomsen}
\affiliation{Q-CTRL, Sydney, NSW Australia \& Los Angeles, CA USA}
\date{\today}


\begin{abstract}
We describe an experimental effort designing and deploying error-robust single-qubit operations using a cloud-based quantum computer and analog-layer programming access.  We design numerically-optimized pulses that implement target operations and exhibit robustness to various error processes including dephasing noise, instabilities in control amplitudes, and crosstalk.  Pulse optimization is performed using a flexible optimization package incorporating a device model and physically-relevant constraints (e.g. bandwidth limits on the transmission lines of the dilution refrigerator housing IBM Quantum hardware).  We present techniques for conversion and calibration of physical Hamiltonian definitions to pulse waveforms programmed via Qiskit Pulse and compare performance against hardware default DRAG pulses on a five-qubit device.  Experimental measurements reveal default DRAG pulses exhibit coherent errors an order of magnitude larger than tabulated randomized-benchmarking measurements; solutions designed to be robust against these errors outperform hardware-default pulses for all qubits across multiple metrics.  Experimental measurements demonstrate performance enhancements up to: $\sim10\times$ single-qubit gate coherent-error reduction; $\sim5\times$ average coherent-error reduction across a five qubit system; $\sim10\times$ increase in calibration window to one week of valid pulse calibration; $\sim12\times$ reduction gate-error variability across qubits and over time; and up to $\sim9\times$ reduction in single-qubit gate error (including crosstalk) in the presence of fully parallelized operations.  Randomized benchmarking reveals error rates for Clifford gates constructed from optimized pulses consistent with tabulated $T_{1}$ limits, and demonstrates a narrowing of the distribution of outcomes over randomizations associated with suppression of coherent-errors.
\end{abstract}



\maketitle


\section{Introduction}

Quantum computers are growing in complexity, capability, and utility across sectors, realizing major scientific accomplishments in the field of advanced computing~\cite{GoogleSupremacy} despite the fact that their performance remains critically limited by hardware imperfections and instabilities~\cite{Preskill_NISQ}.  Superconducting qubits are a leading hardware candidate for the realization of large-scale quantum computers~\cite{DevoretScience2013,WendinIOP2017, KrantzApplPhysRev2019}, and have enabled a variety of cloud-based systems accessible to the public~\cite{Qiskit_code, Rigetti_forest_code}. 

Until recently, cloud access to such quantum computing hardware has been limited to provision of a default gate set~\cite{Qiskit_code, Rigetti_forest_code, GoogleCirc_code, Microsoft_qdev_kit_code} - implemented as fixed, predefined control waveforms - as is appropriate for end users unfamiliar with the details of the underlying physical hardware. IBM Qiskit Pulse~\cite{Qiskit_paper_mckay2018} is an early attempt to provide analog-layer access to the programming of quantum computer hardware over a cloud API. Instead of only defining circuits composed of pre-calibrated operations, analog-layer access gives users the ability to directly program the time-domain modulation of control waveforms that drive quantum logical operations. IBM Qiskit Pulse provides functionality to modulate the local oscillator (LO) frequency, phase, and amplitude envelope of microwave signals with full control over timing and sequencing (we refer to the resulting objects as ``pulses''). In this approach users may define custom operations and even implement new modes of hardware operation beyond the default gate model~\cite{gokhale2020optimized}.   
 
Accessing this form of analog-layer programming enables the incorporation of low-level quantum control as a strategy to improve the performance of quantum computing hardware. Open-loop dynamic error suppression provides a validated approach to deterministically mitigate errors in individual hardware elements and individual qubit operations~\cite{Viola1998, Byrd1995,Gordon2008, Khodjasteh2009dcg, SoareNatPhys2014, wilhelm2020introduction, ball2020software}, complementing the considerably more demanding task of quantum error correction~\cite{Preskill_Layered,Khodjasteh2005, khodjasteh_rigorous_2008, Jones_PRX_2012}. This framework permits the implementation of new error-robust quantum logic operations by exploiting symmetries in the mathematical space of quantum-logic operations and the ubiquitous presence of temporal and spatial noise correlations in real laboratory environments~\cite{edmunds_2020}.

In this manuscript we employ analog-layer programming on superconducting cloud quantum computer hardware to implement and test a new error-robust single-qubit gate set. Analog pulse waveforms are numerically optimized using a custom Tensorflow-based package~\cite{QCTRL_Optimization} to enact gates resilient against dephasing, control-amplitude fluctuations, and crosstalk. We perform experiments on IBM Quantum hardware backends {\sl Valencia} and {\sl Armonk} and program in the IBM Qiskit Pulse API~\cite{Qiskit_PulseAPI}, describing protocols for calibration of the programmable control channels, as well as tuneup of the optimized pulses to account for small distortions.  Our results consistently reveal that pulses optimized to be robust against amplitude or dephasing errors, and including both a $30$-MHz-bandwidth sinc-smoothing-function and temporal discretization to match hardware programming, outperform default calibrated DRAG~\cite{Qiskit_DRAG} operations under native noise conditions.

We use an analytic framework based on previously published error models to quantitatively analyze pulse performance in the presence of both coherent and incoherent errors~\cite{Oliver:2013}. Experimental measurements show good agreement with a model incorporating both coherent rotation errors and an incoherent $T_{1}$-type decay.  Using optimized pulses, extracted measures of the effective-error-per-gate demonstrate improvements up to and exceeding an order of magnitude in single-qubit coherent-error rates, with average device-wide performance improvements $\sim 5\times$. The same optimized pulses reduce gate-error-variability across qubits and over time by an order of magnitude; we demonstrate homogenization of qubit performance in a narrow band around a mean error better than the best-performing default pulse, and extension of the useful calibration window to approximately one week as compared with the daily calibration cycle required for default pulses.  These experiments are cross-validated with Clifford randomized benchmarking where we observe that optimized pulses outperform default pulses by a narrower margin due to the insensitivity of randomized benchmarking to certain coherent errors~\cite{Ball_RB, Proctor_RB}.  Optimized pulses exhibit key signatures of coherent-error-suppression via narrowing of the randomized-benchmarking distribution over sequences. Independent quantitative analyses suggest error-robust optimized pulses exhibit $T_{1}$-limited performance, and extracted incoherent-error magnitudes are consistent with expected errors using tabulated $T_{1}$ times as reported from the hardware backend.  

Finally, we demonstrate that for parallel gate operations (all qubits on a chip driven simultaneously) error-robust optimized pulses exhibit performance largely indistinguishable from serial gate operations (all operations temporally separated such that only one qubit is driven in any timestep), while default pulses are degraded by $\sim 2\times$.  Our measurements - and a change in the observed error-dynamics of the default pulses under parallel gate operation - suggest that under parallel execution, gates on certain qubits exhibit an additional error channel comprising a combination of a direct $\hat{\sigma}_{x}$ coupling as well as an effective AC-Stark shift proportional to $\hat{\sigma}_{z}$. Here, amplitude-noise-robust pulses demonstrate the best performance, providing $8.6\times$ reduction in effective-gate-error, averaged across all five qubits and data sets acquired over a four-day measurement period.

The remainder of this manuscript is organized as follows.  In~\cref{Sec:Optimization} we introduce the physical model on which we perform numeric optimization, and indicate the specific optimization approach employed to build consistency with IBM Quantum hardware constraints.~\cref{Sec: Implementation} describes the process of mapping optimized pulses to executable commands via the Qiskit Pulse API, pulse calibration on hardware, and validation of noise-robustness for optimized pulses. This is followed by an extensive performance comparison under native operating conditions using various benchmarking techniques in~\cref{Sec:Nativenoise}.  Additional benefits in ``virtualizing'' error statistics are demonstrated in~\cref{Sec:ErrorVirtualization} before we present evidence of crosstalk-error mitigation using optimized pulses in~\cref{Sec:Crosstalk}.  We conclude with a brief discussion and future outlook in~\cref{Sec:Conclusion}.

\section{Error-robust pulse optimization for cloud QC hardware\label{Sec:Optimization}}
Our task is to define drop-in replacement definitions for the standard physical gates in use on IBM Quantum hardware, based on the concepts of open-loop control.  We target new solutions which provide high-fidelity operations in the presence of instabilities and noise in the hardware, often referred to as error-robust or dynamically-corrected gates.  Experiments are carried out using a select set of IBM Quantum hardware backends supporting the Qiskit Pulse programming framework.  The relevant hardware systems used in this study were {\sl Armonk}~\cite{Armonk:2020} and {\sl Valencia}~\cite{Valencia:2020}, offering one and five qubits respectively.  All programming was conducted remotely via the Qiskit Python package in combination with custom optimization and transpiler tools from Q-CTRL, again accessed via Python.

\begin{figure}[t]
\centering
\includegraphics[width=\columnwidth]{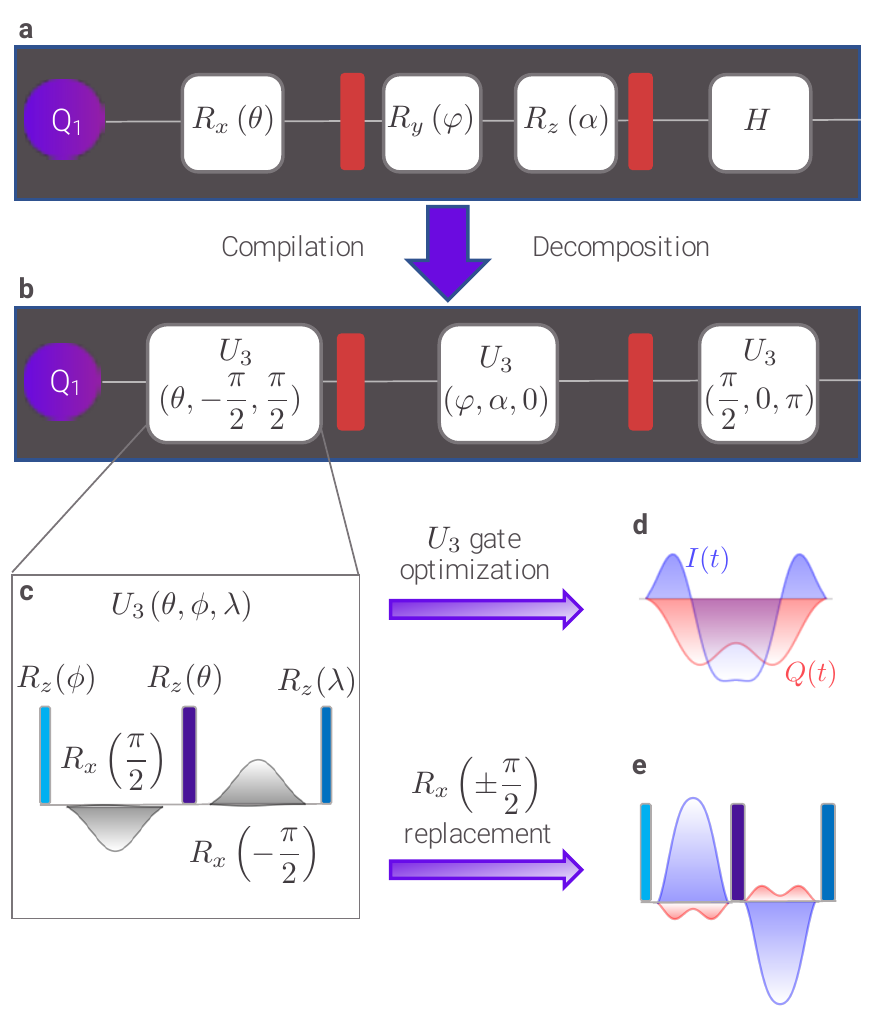}
\caption{ Schematic description of pulse-waveform integration into compiled Qiskit circuits.  (a) A single-qubit circuit composed of physical rotations and ``barriers'' (red bars), that prevent a compiler from mathematically condensing a circuit, implements a desired transformation~\cite{Qiskit_code}.  (b) The Qiskit compiler converts operations into a $U_{3}$ decomposition, respecting barriers established in the circuit. (c) Each $U_{3}$ operation is decomposed into driven rotations, $R_{x}(\pm \pi/2)$, and virtual frame shifts.  (d-e) Two optimization paths are pursued to craft replacement operations with the same net effect.  Here red and blue waveforms represent the Hamiltonian terms $I(t)$ and $Q(t)$ composing the optimized pulse waveform $\gamma(t)$. 
\label{fig:circuit_diagram}}
\end{figure}

Arbitrary single-qubit operations are defined and compiled in Qiskit using an Euler-angle decomposition in the Cartesian basis (\cref{fig:circuit_diagram}) of the form
\begin{align}\label{Eq:U3_default_gate}
U_{3}(\theta, \phi, \lambda) = R_z(\phi) R_{x}\left(-\frac{\pi}{2}\right) R_z(\theta) R_{x}\left(\frac{\pi}{2}\right) R_z(\lambda),
\end{align} 
where $R_z(\xi)$ is a virtual gate (passive frame shift) through phase $\xi$, and $R_{x}(\pm \frac{\pi}{2})$ denotes a driven $\pm \frac{\pi}{2}$ rotation about the $x$-axis. Qiskit Pulse provides flexibility to define optimized replacements for the $R_{x}(\pm\pi/2)$ rotations in~\cref{Eq:U3_default_gate}, or to directly deploy resonantly-driven pulses implementing arbitrary rotations $R_{\hat{\mathbf{n}}}(\theta)$ through angle $\theta$ about an axis $\hat{\mathbf{n}} = (\cos(\phi), \sin(\phi), 0)$ in the $xy$-plane on the Bloch sphere. We demonstrate benefits in performance using optimized pulses within this expanded gate set~\cite{gokhale2020optimized}.

To define the appropriate optimized replacement we employ a model-based approach which begins with defining the total system 
\begin{align}
H_\text{tot}(t) = H_\text{ctrl}(t) +  H_\text{leakage} + H_\text{noise}(t)
\end{align}
\noindent where the single-transmon control Hamiltonian is expressed as
\begin{align}
H_\text{ctrl}(t) & = \frac{1}{2}\left(\Omega(t)e^{i\phi(t)} \hat{a} + \text{H.C.}\right)\\
&= \frac{1}{2}I(t)\hat{a}_\text{I} + \frac{1}{2}Q(t)\hat{a}_\text{Q}.
\end{align}
Here $a$ and $a^{\dagger}$ are the lowering and raising operators for the transmon, and 
we have introduced Hermitian quadrature operators functioning in a manner analogous to the Pauli $\hat{\sigma}_{x,y}$ operators for a pure two level system, defining $\hat{a}_\text{I} =  \hat{a}+\hat{a}^\dagger $ and $\hat{a}_\text{Q} =  -i\left(\hat{a}-\hat{a}^\dagger\right)$.  In the control Hamiltonian, the coupling term $\gamma(t) = \Omega(t) e^{i\phi(t)}\equiv I(t) + i Q(t)$ represents the complex-valued time-dependent (control) pulse waveform.  

Next we define the Hamiltonian terms for the dominant error sources: subspace leakage and fluctuations in either the pulse waveform or ambient magnetic fields,
\begin{align}
H_\text{leakage}(t) & = \frac{1}{2}\chi \left(\hat{a}^2\right)^\dagger \left(\hat{a}^2\right) \\
H_\text{noise}(t) & = H_\text{deph}(t) + H_\text{amp}(t).
\end{align}
Here the noise Hamiltonian is written generically, with contributions from dephasing (deph) and over-rotation, or multiplicative amplitude (amp) errors, taking the form
\begin{align}
H_\text{deph}(t) & = \frac{1}{2}\Delta(t)\hat{a}^{\dagger}\hat{a}\\
H_\text{amp}(t) & = 
\begin{cases}
\epsilon_{\Omega}H_\text{ctrl}(t)  & 
\\
\frac{1}{2}\epsilon_\text{I}I(t)\hat{a}_\text{I} + \frac{1}{2}\epsilon_\text{Q}Q(t)\hat{a}_\text{Q}  
 & 
\end{cases}
\end{align}
where the first case in $H_\text{amp}$ describes common-quadrature noise on the pulse modulus, and the second describes differential noise on the $I/Q$ components. 

Employing error-robust control in this setting involves defining the pulse waveform $\gamma(t)$, to effectively compensate for the presence of both $H_\text{leakage}$ and $H_\text{noise}(t)$ during execution of an arbitrary unitary operation $U_\text{ctrl}(\tau)$ over time $\tau$. This new pulse waveform (\cref{fig:circuit_diagram}) will then be used to implement the target unitary when appropriately compiled in a circuit.  In principle these Hamiltonian terms may have unknown magnitude and temporal dynamics; we require a high-performing solution under a wide range of conditions, making this a problem in robust control rather than optimal control.  Formally, given the Schr{\"o}dinger equation
\begin{align}\label{eq:control_hamiltonian_schrodinger_equation}
    i\dot{U}_\text{ctrl}(t) = H_\text{ctrl}(t) U_\text{ctrl}(t),
\end{align}
one attempts to define $H_\text{ctrl}(t)$ such that
\begin{align}
\label{eq:optimal_control_condition}
U_\text{target} &= U_\text{ctrl}(\tau)\\
\label{eq:robust_control_condition}
\tilde{U}_\text{noise} & = \mathbb{I}
\end{align}
where $\mathbb{I}$ is the identity operation on the control system.  The first line addresses the quality with which the executed unitary matches a target operation.  The second incorporates the distorting effect of noisy dynamics via expression of the total propagator as
\begin{align}\label{eq:total_unitary_as_product_of_ctrl_and_err}
\tilde{U}_\text{noise}(\tau) = U_\text{tot}(\tau) U_\text{ctrl}(\tau)^\dag.
\end{align}
The residual operator $\tilde{U}_\text{noise}(\tau)$ defined in~\cref{eq:total_unitary_as_product_of_ctrl_and_err} is referred to as the \emph{error action operator}. This unitary satisfies the  Schr\"{o}dinger equation in an interaction picture co-rotating with the control. The robust-control problem may then be formalized by the additional constraint
\begin{align}\label{eq:robust_control_condition_2}
\tilde{U}_\text{noise}  = \mathbb{I}.
\end{align}

Control-robustness is evaluated as the noise-averaged operator distance between the error action operator~\cref{eq:total_unitary_as_product_of_ctrl_and_err} and the identity, giving a corresponding fidelity measure 
\begin{align}\label{eq:fidelity_measure_for_robust_control_no_projection}
\mathcal{F}_\text{robust}(\tau) =
\left\langle
\left|
\frac{1}{D}
\Big\langle
\tilde{U}_\text{noise}(\tau), \mathbb{I}
\Big\rangle_{F}
\right|^2
\right\rangle,
\end{align}
where the outer angle brackets $\langle\cdot\rangle$ denote an ensemble average over realizations of the noise processes, and the inner angle brackets $\langle\cdot,\cdot\rangle_{F}$ denotes a Frobenius inner product.  See Reference~\cite{ball2020software} for a detailed discussion of error-robust control creation in arbitrary-dimensional Hilbert spaces.

We design error-robust pulses using a flexible numerical optimization package based on TensorFlow~\cite{ball2020software} allowing the system Hamiltonian to be represented using almost-arbitrary functions of the controllable parameters. Hamiltonian terms defining qubit frequencies and anharmonicities are taken from backend specifications provided for each hardware system for either optimization or implementation. The relevant IBM hardware systems exhibit single-qubit coupling rates to the pulse waveform that are significantly throttled relative to many laboratory systems; the gate time for a default $R_{x}(\pi) \equiv U_{3}(\pi, -\pi/2, \pi/2)$ rotation takes values $\tau_{g}\approx 284$~ns on {\sl Armonk}~\cite{Armonk:2020} and $\tau_{g}\approx 71$~ns on {\sl Valencia}~\cite{Valencia:2020}. Therefore it is generally sufficient to consider a two-level approximation and ignore leakage terms when achieving high-fidelity gates.

Optimization defines a map from controllable parameters to the Hamiltonian; the underlying structure of this map is a TensorFlow graph, which can be efficiently evaluated and differentiated using the TensorFlow automatic differentiation library. Once constructed, this mapping (or graph) is used to calculate the optimization cost function. The resulting optimized pulses thus achieve the desired objectives, and do so within the constraints imposed by the user-defined Hamiltonian structure. Full code-based examples are available from Refs.~\cite{QCTRL_Optimization, QCTRL_Bandlimited}.

The limited hardware-frequency-response of IBM systems is captured directly in our optimizations~\cite{QCTRL_Bandlimited}. Based on our observations of the IBM hardware, we focus on pulses which are transformed as $\gamma(t)\to L\{\gamma\}(t)$, where $L\{\cdot\}$ represents convolution of the time-domain control with the impulse-response of a linear time-invariant (LTI) filter.  We select a sinc function to serve as a smoothing filter which efficiently imposes bandwidth limits on the optimized pulses (see~\cref{fig:allpulses}), before resampling into discrete time for compatibility with Qiskit Pulse programming~\cite{Qiskit_PulseAPI}.  We have also tested pulses optimized using a bound-slew constraint (limiting the per-segment rate of change in the waveform, $\Delta \gamma/\Delta t$), and pulses with completely unconstrained slew rates, however these solutions produced inferior performance compared to smoothed pulses, and we do not present them here. 

The optimized pulses presented here are designed to produce robustness against either dephasing errors, amplitude errors, or both errors simultaneously (dual robust). In general they provide additional robustness against error at the expense of extended operational duration; the optimized pulses used in this work are longer than their IBM default counterparts by factors of $\sim40 - 140\%$ in the {\sl Valencia} backend. Full details of optimized pulses created and used in this work are presented in~\cref{fig:allpulses} in~\cref{app:valencia_pulses}.

\section{Implementation and Validation of Error-Robust Optimized Pulses\label{Sec: Implementation}}
\begin{figure*}[tp!]
\centering
\includegraphics[width=\textwidth]{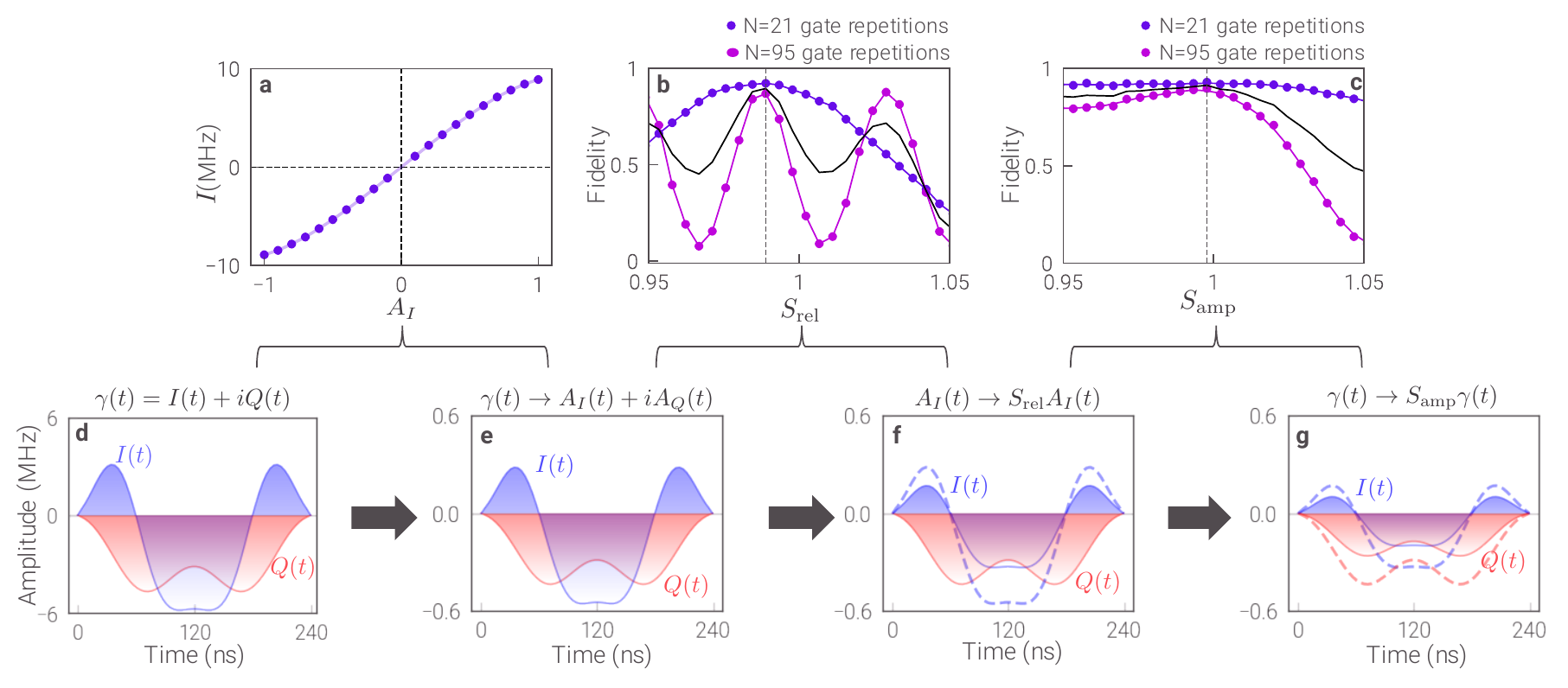}
\caption{Pulse transformation from Hamiltonian representation and calibration for execution on hardware.  (a) Response of {\sl Armonk} hardware defining the mapping $(I,Q)\rightarrow (A_I,A_Q)$ from physical to programmable waveforms, used to perform the calibration step in panels (d)-(e).  Each point in panel (a) corresponds to a Rabi rate derived from fitting to a standard Rabi flopping experiment at fixed square-pulse amplitude of varying duration.  (b) Calibration measurements used to determine the calibration parameter $S_\text{rel}$ used to tune the relative scaling between $I/Q$ waveforms, resulting in the deformation (not to scale) in panels (e)-(f).  Each trace in panel (b) corresponds to a measured state fidelity for different pulse-repetition numbers.  Finding a common maximum denoted by the average (solid line) provides enhanced sensitivity to $S_\text{rel}$ while excluding net rotation errors $\mod(2\pi)$.  (c) Similar procedure used to find the calibration parameter $S_\text{amp}$ used to tune the common-quadrature scaling of the waveform, resulting in the deformation (not to scale) in panels (f)-(g).   
\label{fig:calibration}}
\end{figure*}

Qiskit Pulse formatting permits the definition of analog pulse waveforms corresponding to programming of the $I$ and $Q$ terms in our control Hamiltonian.  IBM default pulses are implemented by simply calling their daily calibrated $U_{3}(\theta, \phi, \lambda)$ pulses with the appropriate angles.

For optimized controls, one could opt to replace the individual $R_x(\pm \pi/2)$ pulses in the default $U_{3}$ decomposition with corresponding optimized pulses or, alternatively, replace the entire $U_{3}$ construction with an optimized pulse implementing the same gate, as shown in~\cref{fig:circuit_diagram}. All pulses in this work were obtained using the latter approach. Appropriate for current hardware settings, optimized waveforms are discretized in units of the minimum backend timing resolution, $dt = 0.22$ ns, and are designed to have total duration (gate time) $\tau_{g} = n_{1}\tau_{s}$ consisting of $n_{1}$ segments of uniform duration $\tau_{s} = n_{2} dt$, subject to the condition that $n_{1}n_{2} = 16 m$ for integers $n_{1}, n_{2}, m$. Once optimized, as described in~\cref{Sec:Optimization}, execution of a custom pulse in Qiskit requires only a single command \texttt{Play(Waveform(custom\_pulse), drive\_channel)}.

Arbitrary multiqubit circuits combining multiple default or optimized-pulse waveforms in time across all available qubits are sequenced into a Qiskit Pulse compatible format via a custom transpiler.  Circuits such as randomized benchmarking, calibration routines, or algorithms, may be efficiently deployed replacing effective gates with arbitrary pre-defined pulse waveforms.  See Ref.~\cite{QCTRL_IBMSingleQubit} for code-based implementation using Qiskit Pulse.

The first step of pulse execution in hardware requires mapping a complex-valued and piecewise-constant waveform $\gamma(t)$ -- the result returned from a numerical optimization -- to instructions for the Qiskit Pulse API (\cref{fig:calibration}). Each analog pulse is defined for IBM Quantum hardware in terms of dimensionless variables $A = A_I + i A_Q$ with real and imaginary components proportional to physical input voltages, and bounded as $A_{I,Q} \in [-1,1],\;\;|A|\le 1$. A physical calibration routine is used to establish the map $(A_I, A_Q) \leftrightarrow (I, Q)$, from programmable amplitudes to control-pulse amplitudes in the laboratory frame.  We first coarsely calibrate the channels $(A_I, A_Q)$, by measuring Rabi oscillations in the time-domain using square, fixed-amplitude, single-quadrature pulses on $I$ or $Q$ channels independently.  Fitting the respective Rabi oscillations to a cosine, we extract the corresponding Rabi frequency and interpolate between measurements to produce a smooth functional dependence $(A_I, A_Q) \leftrightarrow (I, Q)$, as shown in~\cref{fig:calibration}a.  A desired waveform may then be predistorted via this interpolation function to account for any nonlinearities in the hardware response.  

Additional hardware imperfections are routinely experienced leading to small offsets and distortions in the pulse waveforms programmed by the variables $A_{I/Q}(t)$. We implement fine calibration of the absolute and relative scaling of the $I$ and $Q$ channels by expressing the pulse waveform using two scaling factors as $\gamma(t) = S_\text{amp} (S_\text{rel} A_{I}(t) + i A_{Q}(t))$.  Calibration proceeds by applying repeated pulses to amplify coherent miscalibration errors and varying the parameters $S_\text{amp}$ and $S_\text{rel}$ for different numbers of pulse repetition (\cref{fig:calibration}b, c). Scaling factors are selected which simultaneously yield maximum state-transfer fidelity for two different calibration-sequence lengths.  Supplementary calibration sequences which maximize sensitivity to the orthogonality of $R_{x}(\theta)$ and $R_{y}(\theta)$ may also be employed and follow a similar routine.

 \begin{figure}[t!]
\centering
\includegraphics[width=\columnwidth]{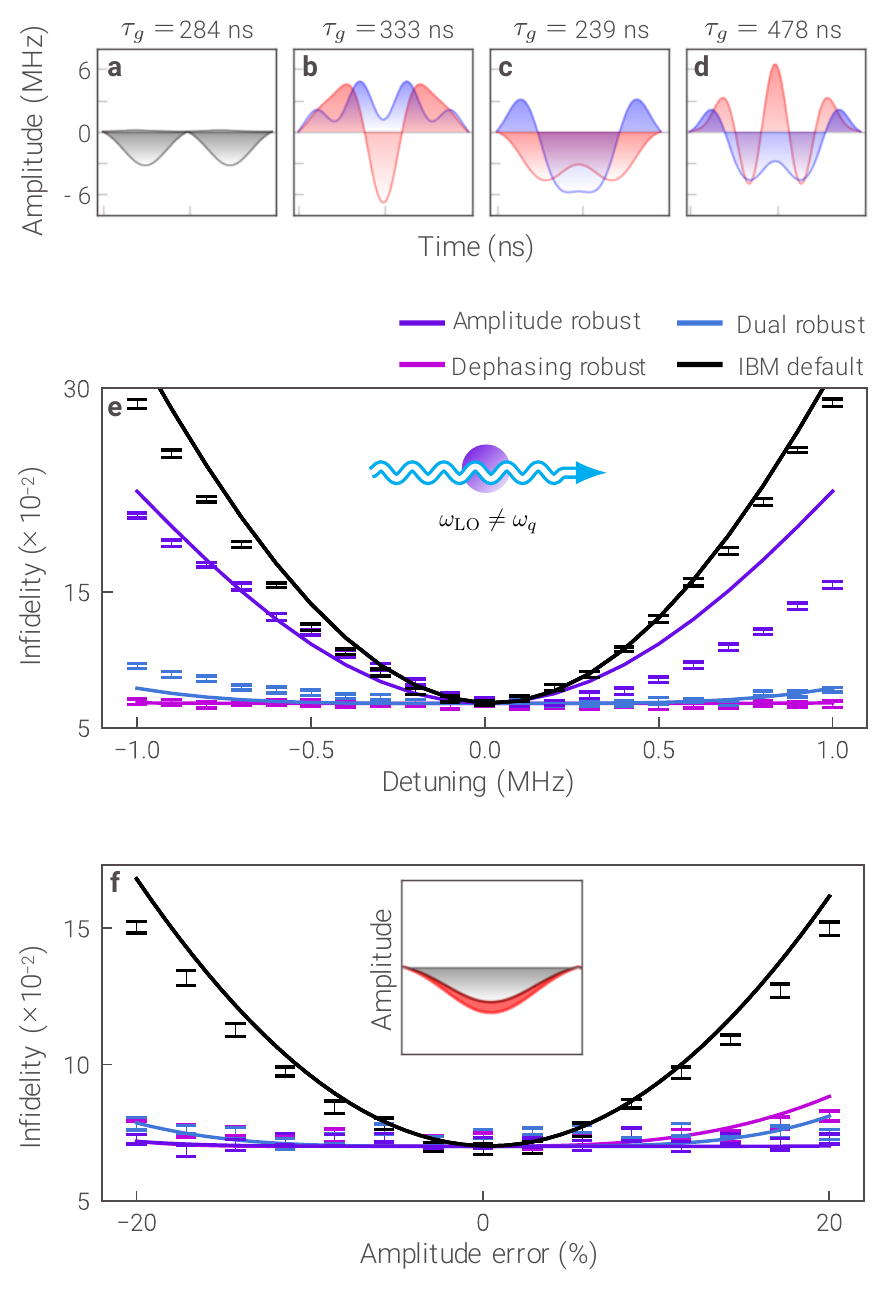}
\caption{Demonstration of optimized-pulse robustness against dominant error sources, using the single-qubit backend {\sl Armonk}. (a) IBM default pulse, (b) amplitude-robust pulse, (c) dephasing-robust pulse (d) dual-robust pulse. All pulses implement a $R_{x}(\pi)$ gate.  Red and Blue represent $I(t)$ and $Q(t)$ respectively.  On this backend Rabi rates are reduced relative to other hardware, resulting in $\sim2-3\times$ longer pulses than executed on {\sl Valencia}. (e) Infidelity for each pulse identified above as a function of an engineered detuning $\Delta = (\omega_{q} - \omega_\text{LO})/2\pi$, obtained by sweeping the LO frequency, $\omega_\text{LO}$, away from the qubit resonance, $\omega_{q}$, determined by the IBM backend. Lines represent theoretical simulations and the error bars represent the standard deviation over 20 independent experimental measurements for each pulse type. Engineered detunings are established in the Qiskit Pulse API. The solutions from simulations are vertically offset to match the baseline measurement-limited experimental fidelity.  (f) Similar as above for a pulse-amplitude error. Inset shows schematic change in pulse area leading to an over/under-rotation error due to the applied pulse-amplitude error.
\label{fig:robustness_verification}}
\end{figure}

Once properly calibrated we can begin to explore the performance enhancements delivered through use of optimized pulses.  We begin by comparing the error-robustness of optimized, calibrated $R_{x}(\pi)$ pulses to the IBM default implementation on IBM's {\sl Armonk} system (\cref{fig:robustness_verification}a-d).  Here we apply an intentionally engineered detuning between the microwave frequency and qubit resonance (\cref{fig:robustness_verification}e), or an over-rotation error realizing quasistatic terms in $H_{\text{noise}}$ (\cref{fig:robustness_verification}f). As the strength of the detuning or over-rotation error is increased the state-transfer fidelity of the IBM default pulse degrades rapidly. By contrast, noise processes corresponding to up to $1$-MHz detuning from the qubit frequency and deviations of 20 \% in pulse amplitude, we see that optimized robust pulses provide high-fidelity state transfer, a key signature of  robustness to the engineered $H_{\text{noise}}$ term.  This robustness is primarily exhibited against the noise channel for which optimization was performed, though the dephasing-robust pulse also shows some robustness against amplitude errors (see also~\cref{fig:allpulses} in~\cref{app:valencia_pulses} for information on the robustness of pulses designed for operation on {\sl Valencia}).  Data match theoretical predictions calculated based on the implemented waveforms well for all pulses, with minor deviations observed for large error magnitudes.   We note that optimized pulses lacking band limits were not able to demonstrate this form of robustness, likely due to low-fidelity reproduction of the pulse waveform at the qubits.

These initial measurements validate that we are able to implement calibrated operations enacting high-fidelity state-transfer to within measurement error, and that optimized pulses exhibit robustness against the appropriate Hamiltonian terms. Next we will provide a comparative performance analysis for these pulses against the calibrated IBM default pulses under native operating conditions on the five-qubit {\sl Valencia} system.

\section{Error mitigation under native hardware performance conditions\label{Sec:Nativenoise}}

Under native operating conditions single-qubit gates are tabulated to achieve operational fidelities exceeding $99.9\%$. By contrast, for the backends used in our experiments, the reported readout error using Qiskit ``measurement level 2''~\cite{Qiskit_Ignis} without any post-processing can exceed $5\%$, as seen in~\cref{fig:robustness_verification}. To ensure we are able to perform detailed comparisons between pulses beyond the limits imposed by so-called state-preparation and measurement (SPAM) errors, we employ a range of ``error-amplifying'' protocols which are SPAM-insensitive; using these it is possible to extract an effective error rate much smaller than the SPAM threshold.

\begin{figure}[tbh]
\centering
\includegraphics[width=\columnwidth]{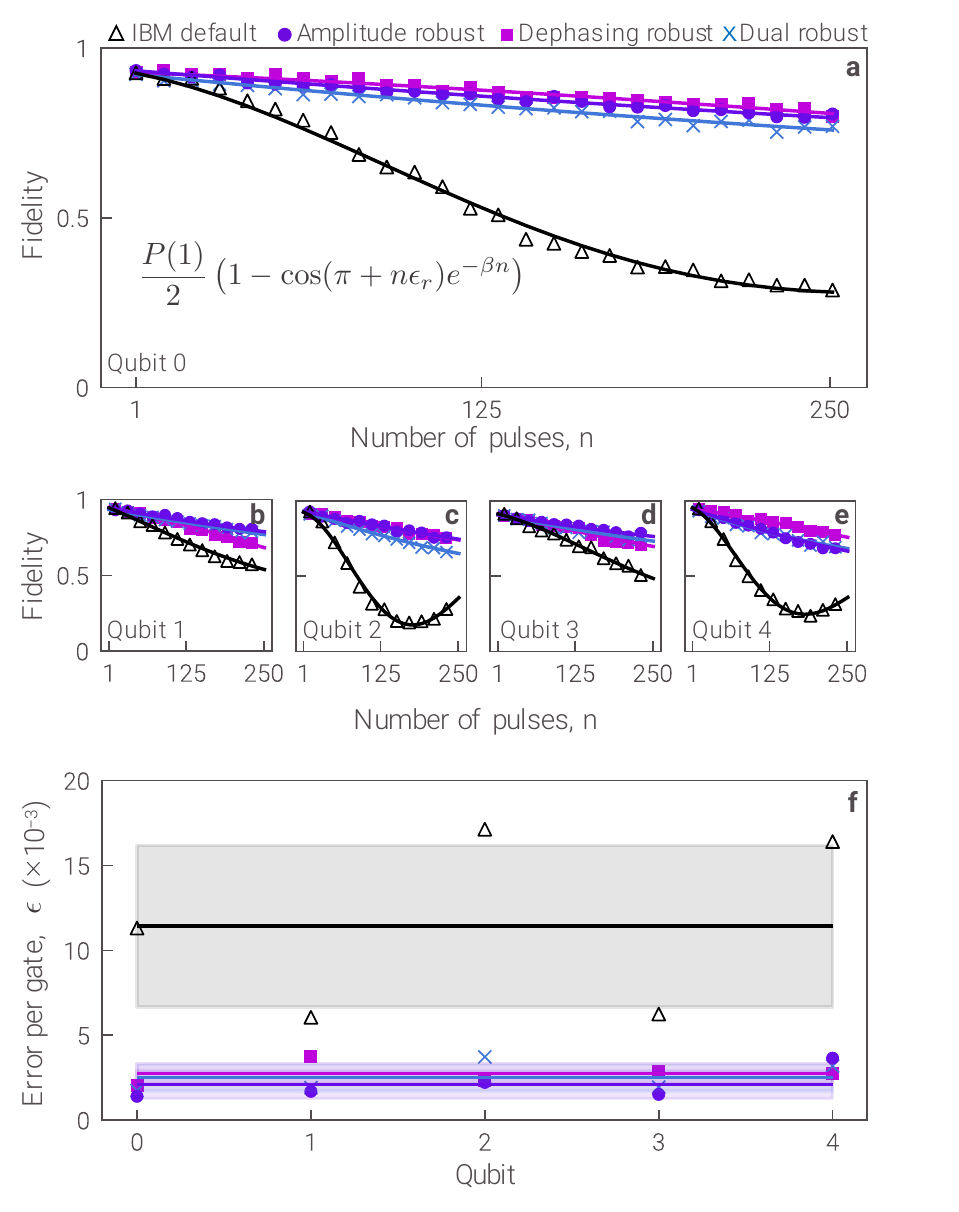}
\caption{Gate performance for all five qubits on the IBM {\sl Valencia} device on August 5, 2020.  Full details of the pulses in use and their durations are presented in~\cref{fig:allpulses}, but operations are approximately $\sim2-3\times$ shorter than on {\sl Armonk}.  (a-e) Single-qubit data sets for repeated pulses applied independently to each qubit, overlaid with theoretical fits. (f) The effective error rate extracted from the fit parameters, $\epsilon$, is shown for four different pulses: IBM default $U_{3}$ (triangles), amplitude-robust (circles), dephasing-robust (squares), and dual-robust (crosses). Solid lines represent the average across the device, while the shaded region  corresponds to the average $\pm$ the standard deviation, $\sigma_{q}$. All robust pulses outperform the default for every qubit.
\label{fig:single_day_performance}}
\end{figure}

The first sequence we employ consists of repeated application of a target rotation.  This sequence is maximally sensitive to coherent over or under rotations arising from various noise sources. All repetitions are constructed to comprise an odd-integer multiple of $R_{x}(\pi)$.  As the repetition number grows, we observe a gradual deviation of the ideal state-transfer probability to the excited state which differs substantially between different pulse implementations (\cref{fig:single_day_performance}a).  Deviation from ideal state transfer is slowest for optimized pulses designed to suppress terms in $H_{\text{noise}}$, and worst for the default $U_{3}$ gate, across all qubits (\cref{fig:single_day_performance}a-e).  This indicates that the numerically-optimized pulses are more robust to naturally occurring noise, drifts, and miscalibrations in either pulse amplitude, duration, or frequency detuning.

Data are fit (\cref{fig:single_day_performance}a) with a decay curve modelling simultaneous coherent rotation errors and incoherent $T_{1}$-type decay, in line with published error models~\cite{Oliver:2013}
\begin{align}\label{eq:expcosdecay}
    \frac{P(1)}{2}\left(1-\cos(\pi+n\epsilon_{r})e^{-\beta n}\right).
\end{align}
\noindent Here, $P(1)$ corresponds to the value of the first point in the data set and represents a constant (but variable between data sets) offset associated with measurement error.  We allow both the angular rotation error, $\epsilon_{r}$, and the population-decay error per gate, $\beta$, to serve as fit parameters and observe excellent agreement with data.  The total effective gate error, $\epsilon$ is determined from the quadrature sum of coherent and incoherent error terms, $  \epsilon = \sqrt{\epsilon_{r}^2 +\beta^2}$.  

Typical values of $\epsilon$ are in the range $10^{-3}-10^{-2}$, somewhat larger than tabulated randomized-benchmarking results (see~\cref{Sec:ErrorVirtualization} for full discussion).  The extracted incoherent decay is also somewhat stronger than the backend tabulated values reported by IBM for the default pulses.  By contrast, the extracted values of $T_{1} \sim \tau_{g}/\beta$ for optimized pulses on each qubit compare well against tabulated values, and exhibit fluctuations between data sets comparable to the statistical deviation recorded provided by IBM's internal calibration procedures (see~\cref{fig:T1} in~\cref{app:T1}). Alternate fitting approaches, such as fixing all values of $P(1)$ across a device, or fixing $T_{1}$ based on measured values visibly degrade fit quality, but do not materially alter the best-fit value of $\epsilon$. We note that qubits two and four routinely show substantial deviation in performance from tabulated values (possibly due to transient two-level fluctuators~\cite{PhysRevLett.121.090502, PhysRevLett.123.190502}).

The effective gate error, $\epsilon$, for multiple optimized pulses extracted from data collected within a single-day calibration period on the IBM Quantum device is shown in~\cref{fig:single_day_performance}f.  Across all qubits we observe superior performance from numerically optimized robust pulses, with marginal difference between them on this day. In these representative data the observed infidelity for numerically-optimized pulses is dominated by the effective exponential decay which we attribute to $T_{1}$.  Quantitatively, in this single-day data set we observe up to $8\times$ improvement in the reduction of single-qubit $\epsilon$.  Similar results were achieved over a six-day measurement window Aug 4-9, 2020 shown in~\cref{fig:device_wide_performance_sequential}, and over an additional window in the range May 18-22, 2020.  We observed a maximum of $10.8\times$ improvement in single-qubit $\epsilon$.

\section{Reducing space-time error fluctuations and correlations\label{Sec:ErrorVirtualization}}

\begin{figure*}[tp!]
\includegraphics[width=\textwidth]{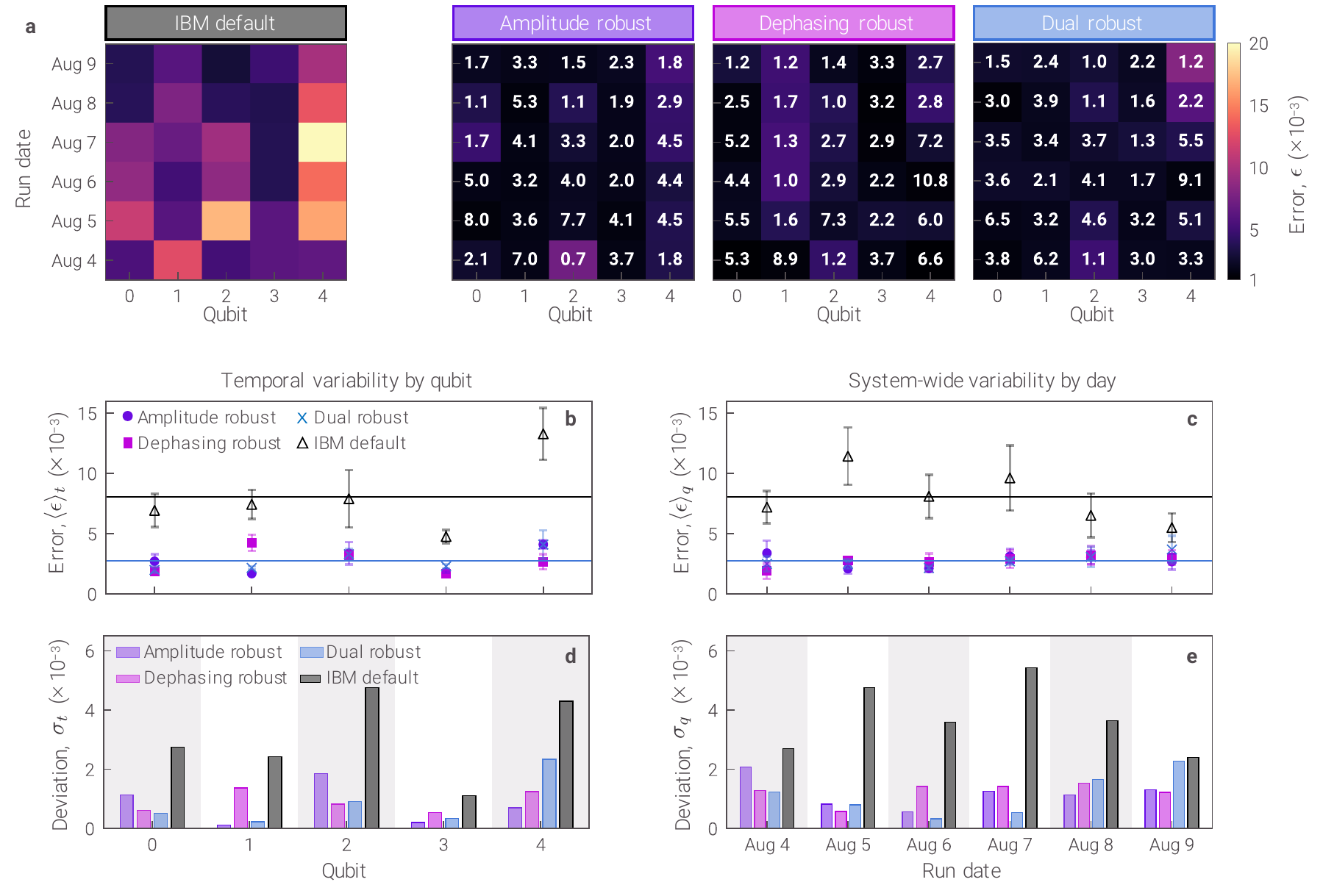}
\caption{(a) Effective error, $\epsilon$, across qubits and measurement days expressed as a color scale. All experiments with robust pulses were carried using a calibration performed on August 3, while the left plot used the IBM default pulse from the most recent daily hardware calibration. The gate improvement from robust pulses for each qubit and day is quantified by the error reduction, ($\epsilon_{\rm default}/\epsilon_{\rm robust}$), given explicitly by the numbers overlaid on the colorplot. The temporal and system-wide variabilities are summarized, respectively, by panels (b, d), and (c, e). The effective error averaged over different days, $\langle \epsilon \rangle_t$, for each qubit is shown in (b), while (c) displays the average over all qubits in the device, $\langle \epsilon \rangle_q$, for different days. The standard deviations for each average error are shown as error bars in (b, c) and in the bar plots (d, e) for ease of comparison between the different pulses. All robust pulses strongly reduce both temporal and system-wide variabilities.
\label{fig:device_wide_performance_sequential}}
\end{figure*}

IBM Quantum hardware exhibits natural performance variation across qubits on a device and over time. For instance, the tabulated $T_{1}$ and $T_{2}$ values show typical daily variability of several tens of percent and up to over $2\times$ for individual qubits~\cite{murali2019noiseadaptive}.  Variation of a similar scale is also observed between qubits on a single device.  This spatio-temporal inhomogeneity is reflected in the extracted $\epsilon$ for the IBM default DRAG pulse, displayed as a color scale for the different pulses, across both qubits on a device and measurement days (\cref{fig:device_wide_performance_sequential}a).  For the August 2020 data acquisition period shown here we observe more than $8\times$ difference in $\epsilon_{r}$ from the best to the worst recorded single-device performance (\cref{fig:device_wide_performance_sequential}a). 

This hardware variability is suppressed through use of numerically-optimized pulses. Beginning with~\cref{fig:single_day_performance}f, we observe that the system-wide variability on {\sl Valencia} for each pulse, expressed as the standard deviation of $\epsilon$ across qubits, $\sigma_{q}$, is reduced for all of the optimized error-robust optimized pulses (indicated by colored shading).  For this day, the dephasing-robust pulse provides an $8.2\times$ reduction of $\sigma_{q}$ relative to the IBM default pulses. A similar reduction in performance variability across devices is observed on multiple measurement days.  A more explicit quantitative comparison of the system-wide variability for IBM default and robust pulses is shown in panels (c) and (e) of~\cref{fig:device_wide_performance_sequential}, where the effective error averaged over all qubits (c), $\langle \epsilon \rangle_q$, and the corresponding standard deviations (e) are presented for the different days of data acquisition. For all optimized pulses the variability across devices is reduced; daily average ``error homogenization'' ranges from $\sim1.1-10.8\times$ across this week.

Temporal variation in gate performance for each qubit is also suppressed through use of optimized pulses. Improvement on individual-qubit stability over time using optimized error-robust pulses is summarized in~\cref{fig:device_wide_performance_sequential}(b) and~\cref{fig:device_wide_performance_sequential}(d), which show, respectively, the time-averaged errors, $\langle \epsilon \rangle_t$, for each qubit and their deviations, $\sigma_t$. These plots enable direct comparison of the temporal variability for different pulses on each qubit, with robust pulses reducing the deviation in the time-averaged error from $1.8\times$ (Qubit 1, dephasing-robust) to over $20\times$ (Qubit 1, amplitude-robust) as compared to the IBM default pulse.  Full tabulation of all extracted errors and calculated statistical moments for each pulse type are presented in \cref{app:data_sets}.

These measurements highlight an additional practical benefit; in the data displayed in~\cref{fig:device_wide_performance_sequential}, the IBM default pulses are recalibrated daily, while the numerically-optimized pulses are calibrated only once on 3 August and used with the same setting over the course of the week. Measurements of device performance within a single 24 hour period indicate that calibrations of default pulses begin to go stale after $\sim12$ hours, while the calibration for numerically optimized robust pulses is extended by over $10\times$ to at least six days. Even at the end of the six-day measurement period, each individual optimized pulse type on each qubit still outperforms the default pulse (\cref{fig:device_wide_performance_sequential}(a)). 

\begin{figure}[!]
\centering
\includegraphics[width=\columnwidth]{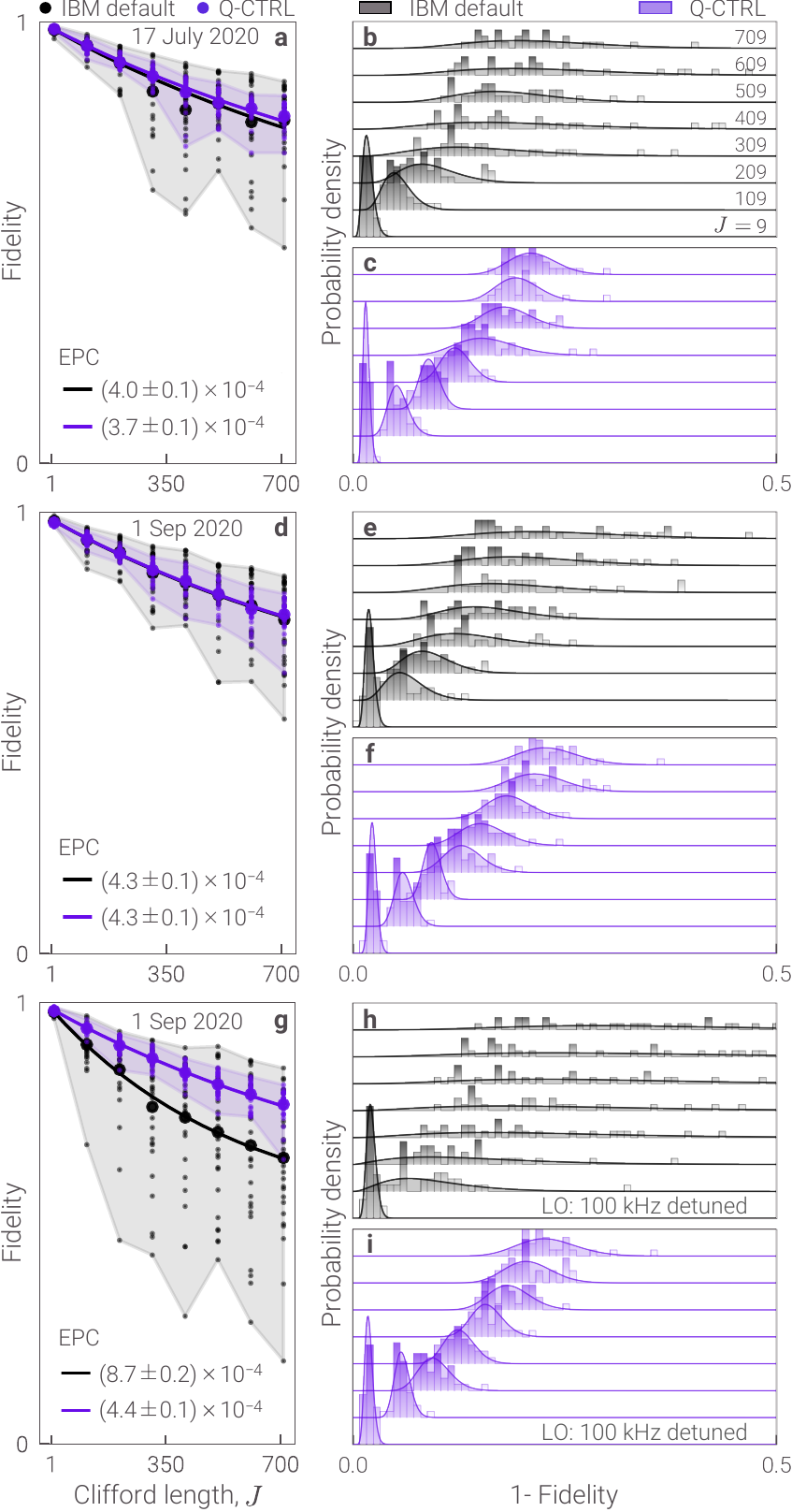}
\caption{Randomized benchmarking comparison between IBM default pulses and optimized dephasing-robust pulses. In these experiments all 24 single-qubit Clifford gates are transpiled using only $R_{x}(\frac{\pi}{2})$ and $R_{x}(\pi)$ driven pulses, interleaved with virtual $R_{z}$ gates as needed. For the IBM default case (gray), $R_{x}(\pi) = R_{x}(\frac{\pi}{2})R_{x}(\frac{\pi}{2})$ where $R_{x}(\frac{\pi}{2})$ is a DRAG pulse as shown in~\cref{fig:allpulses}b. For the Q-CTRL case (purple), $R_{x}(\frac{\pi}{2}, \pi)$ gates are independently optimized pulses as shown in \cref{fig:allpulses}(f,j) respectively. Panel (b) shows histograms of measured fidelity for each sequence length, $J$, using IBM default pulses, overlaid with fits to gamma distributions.  Large variance and skew in distributions for default pulses is indicative of coherent errors~\cite{Ball_RB, Mavadia_RB, edmunds_2020}.  Panel (c) reveal both the narrower distribution and shape better approximated by a Gaussian, indicating decorrelation of coherent errors~\cite{edmunds_2020}.(d-f) shows similar data, repeating the experiment on a later date.  (g-i) show similar data, but imposing a 100-kHz detuning on the LO to reveal the impact of a deliberate (large) frequency miscalibration error, changing the extracted error-per-Clifford (EPC) $p_{RB}$ by $\sim2\times$.}
\label{fig:RB_main_dephasing_pulses}
\end{figure}

Single-qubit measurements have been cross-referenced against Clifford randomized-benchmarking performed using either $U_{3}$ gate constructions composed from default $R_{x}(\pi/2)$ pulses, or combinations of error-robust $R_{x}(\pi)$ and $R_{x}(\pi/2)$ pulses.  We observe that numerically optimized error-robust pulses can outperform the default pulses in the returned $p_{RB}$ despite the extended gate-duration relative to the default, but now by a smaller margin of approximately $10-20\%$ (\cref{fig:RB_main_dephasing_pulses}a), relative to the experiments presented above. In general we find that the extracted $p_{RB}$ for both gate classes vary by date (\cref{fig:RB_main_dephasing_pulses}a,d), and that differences in performance fall within statistical variations between data sets.  In addition, we have observed that the best results are achieved using a compilation involving both driven, optimized $R_{x}(\pi)$ and $R_{x}(\pi/2)$ pulses, rather than constructing all gates from the $U_{3}$ decomposition.  We believe this is due to the reduced execution time and the presence of limiting $T_{1}$ processes (combining two optimized $R_{x}(\pi/2)$ pulses requires $\sim6\%$ longer than a single optimized $R_{x}(\pi)$ pulse).  

The extracted values of $p_{RB}$ are approximately eight times smaller for optimized pulses and nearly $20\times$ smaller for default pulses than the direct probing experiments presented in previous figures.  This discrepancy is consistent with the previously reported reduced sensitivity of randomized benchmarking to coherent errors arising from {\em e.g.} calibration errors and drifts~\cite{Proctor_RB}.  This insensitivity can lead to large deviations between expected and actual hardware performance at the algorithmic level~\cite{Proctor:2020}. Our measurements nonetheless reveal the action of robust optimized pulses in modifying the characteristics of gate-errors through suppression of error correlations.  

The distribution of randomized benchmarking outcomes over sequences contains a key signature of coherent errors in the form of a skewed and broadened distribution~\cite{Ball_RB, Mavadia_RB, edmunds_2020}. For the IBM default pulses this skew and broad variance is apparent, showing a large spread of measured return probabilities over different measured sequences, as illustrated in~\cref{fig:RB_main_dephasing_pulses}a.  By contrast, the numerically-optimized pulses show a narrow and symmetric distribution around the mean.  

The form of these distributions is quantitatively matched to a gamma distribution with a shape varying with the form of the error model, ranging from approximately Gaussian (incoherent errors) to approximately log-normal (coherent errors)~\cite{Ball_RB, Mavadia_RB}.  In~\cref{fig:RB_main_dephasing_pulses}b, c,  histograms of measured infidelities over different randomized sequences are overlaid with fits to a gamma distributions.  Agreement is good in both cases, and highlights the differences between default and optimized pulses.  These observations are consistent with the interpretation that error-robust optimized pulses suppress coherent errors, leaving residual errors dominated by incoherent $T_{1}$ processes, while default pulses remain susceptible to both error processes.  We consider this a form of ``error virtualization'' in which the statistical characteristics of the gates change, even if the observed $p_{RB}$ is only minimally affected~\cite{edmunds_2020}.

Additional experiments performed with pulses designed to suppress both coherent error processes and leakage out of the qubit subspace show similar performance, with differences laying within the statistical variations in the IBM hardware.  The range of randomized-benchmarking studies and comparisons we have undertaken suggests that performance may be limited by Hamiltonian terms which are not accounted for either in published system documentation or associated numerical optimizations, in addition to incoherent $T_{1}$ processes.  

\section{Suppression of crosstalk errors\label{Sec:Crosstalk}}
\begin{figure*}[htb]
\includegraphics[width=\textwidth]{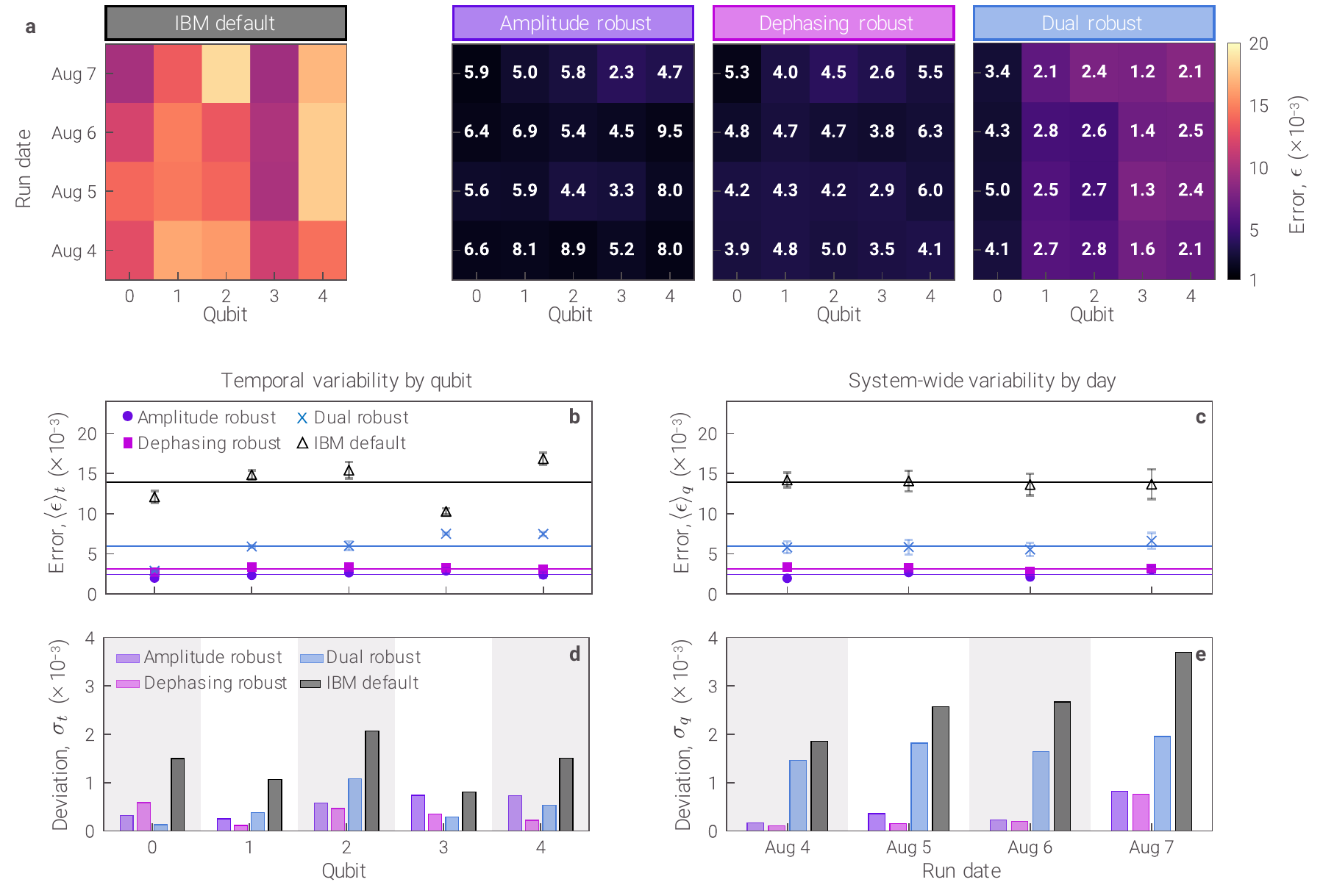}
\caption{Device-wide performance, as in~\cref{fig:device_wide_performance_sequential}, but now for parallelized operations (pulses applied simultaneously on all five qubits). Comparison between panels (a) in this figure and in~\cref{fig:device_wide_performance_sequential} shows the deleterious effect of crosstalk when the IBM default pulses are applied in parallel. In contrast, dephasing- and amplitude-robust pulses perform as well, or even slightly better, than in the serial pulse case. The standard deviations in (d) and (e) show that robust pulses can suppress temporal variability by more than $11\times$ and system-wide variations by more than $17\times$. 
\label{fig:device_wide_performance_parallel}}
\end{figure*}
The measurements presented thus far focus on executing pulses on each qubit sequentially, keeping the remaining qubits idle. This matches the serialization approach taken in standard Qiskit compilers~\cite{Qiskit_Terra}, but misses an opportunity to reduce circuit-execution time and the associated incoherent errors. Understandably, parallel execution is not favored in general as crosstalk errors degrade performance relative to tabulated single-gate metrics without further mitigation strategies~\cite{ding2020systematic}.

We test the benefits of numerically-optimized error-robust pulse design in the presence of crosstalk error by studying gate performance under full parallel execution across all five qubits on {\sl Valencia}~\cite{Valencia:2020}. Again, we measure deviation of a sequence of repeated pulses from unity fidelity, and extract an effective error-per-gate. In this case the best fit to the experimental data using IBM default pulses is consistent with the assumption of a dephasing noise Hamiltonian, leading to a fit function
\begin{align}\label{eq:Gausscosdecay}
    \frac{P(1)}{2}\left(1-e^{-\left(\beta^{\parallel} n \right)^2}\cos(\pi+n\epsilon^{\parallel}_{r})\right).
\end{align}
Fit quality for the default pulses is improved by appropriate selection of the functional form, now including a Gaussian decay, but quantitative measures for effective rotation error, here captured through $\epsilon^{\parallel}_{r}$ are minimally affected. By contrast, we continue to find that best-fit performance for optimized error-robust pulses is realized using~\cref{eq:expcosdecay}. 

Extracting an effective error-per-gate, $\epsilon$, we first observe that performance of the default pulse averaged across the device has degraded by an average of approximately $2\times$ relative to serial execution (\cref{fig:device_wide_performance_parallel}a). Numerically-optimized pulses in parallel execution continue to outperform the default pulse, and even outperform the default pulse using serial execution. Surprisingly, numerically-optimized pulses designed to exhibit robustness to both error quadratures performs the worst among the optimized solutions. Under parallel execution, amplitude-robust pulses marginally outperform dephasing-robust pulses across all days and both in device-wide averages and individual-qubit performance, showing up to $9.5\times$ improvement in reduction of $\epsilon$ for an individual qubit. System-wide and temporally averaged performance for both the dephasing-robust and amplitude-robust pulses is unchanged or modestly improved (consistent with daily experimental fluctuations) relative to serial pulse application (\cref{fig:device_wide_performance_parallel}b-c).
 
Examining the effective incoherent decay constants, $\beta$ and $\beta^{\parallel}$, yields results consistent with the appearance of an additional decay process beyond $T_{1}$ under parallel pulse application. Despite good fit agreement, conversion of the decay constants for the (poorly performing) default and dual-robust pulses to a time constant now substantially underestimates $T_{1}$ relative to tabulated values (see~\cref{fig:T1} in~\cref{app:T1}). By contrast, the dephasing- and amplitude-robust pulses, which perform well, exhibit extracted values of $\beta$ that still agree approximately with the tabulated $T_{1}$ from the hardware backend.  

These performance differences are striking when we examine spatio-temporal variability in gate performance. In addition to overall error-reduction, relative to the default pulse we see up to an average of $\sim \!\! 8.6\times$ reduction of variance across all measurement runs for the best-performing amplitude-robust pulses (\cref{fig:device_wide_performance_parallel}d), and up to $\sim \!\! 12\times$ reduction of variance across qubits on a single day (\cref{fig:device_wide_performance_parallel}e). The absolute magnitude of the temporal and system-wide variability appears somewhat smaller than in serial-execution for these pulses (\cref{fig:device_wide_performance_parallel}b-e), which we believe arises due to crosstalk becoming a single dominant error process. We emphasize that these data sets were acquired in an interleaved fashion with the data in~\cref{fig:device_wide_performance_sequential}, over the window of Aug 4-7, 2020.  

From these data sets we surmise that crosstalk in these devices likely occurs as both a term in $H_{\text{noise}}\propto{\hat{\sigma}_{x}}$ contributing to an effective amplitude error, and an additive term $H_{\text{noise}}\propto{\hat{\sigma}_{z}}$ due to what appears to be an effective AC-Stark shift.  We draw this conclusion for the following reasons;  First, the superior performance of amplitude-robust vs dephasing-robust optimized pulses, and the device-wide performance degradation of the dephasing-robust pulses suggests that the error process induced by parallel operation is not pure dephasing.  Next, the difference in best-fit form for the Gaussian decay of default pulses (while the dephasing-robust and amplitude-robust pulses maintain a simple exponential decay form) suggests the presence of a new quasistatic dephasing term that is suppressed by the robust pulses until their performance is limited by an incoherent error process.  We do not have a full explanation for the poor performance of the``dual-robust'' pulses, but note they are $170$ ns in duration, $70\%$ longer than the dephasing-robust pulses, providing an additional potential source of error. 

\section{Conclusion\label{Sec:Conclusion}}
In this work we have employed the Qiskit Pulse programming API to experimentally test numerically-optimized error-robust pulses implementing single-qubit gates on IBM cloud quantum computers.  We have performed a detailed comparison of different pulse designs incorporating various optimization constraints and objectives.  We find that in realistic circuits~\cite{Proctor:2020}, coherent errors dominate default-pulse performance and are manifested as circuit errors which can be an order of magnitude larger than randomized benchmarking results would suggest.  These errors are suppressed by up to $\sim10\times$ for an individual qubit through use of optimized error-robust pulses under both serial and parallel execution on the five-qubit {\sl Valencia}~\cite{Valencia:2020} hardware backend.  Our results reveal that the best performing single-qubit pulses were those designed to be robust against either control-amplitude or dephasing noise, and incorporating a $\sinc$ smoothing function to account for band-limits on transmission.  Our investigations further suggested the presence of crosstalk errors due to simultaneous dephasing and coherent rotation errors, both of which are suppressed through appropriate choice of robust pulses.

In addition we demonstrate that optimized robust pulses virtualize errors by changing their statistical properties in addition to error magnitudes~\cite{edmunds_2020}.  We use error-robust optimized pulses to demonstrate mitigation of hardware variability across qubits and time, reducing the standard deviation of measured qubit errors by $\sim8.6\times$ over a four-day measurement window.  During this window the robust pulses were never recalibrated, while IBM default pulses were observe to exhibit performance variability on a $\sim12$ hour timescale and require daily recalibration; this corresponds to a net improvement in the calibration window by over $\sim10\times$. Additional randomized benchmarking measurements reveal key signatures of coherent-error suppression using optimized pulses, even as returned values of $p_{RB}$ appear consistent with $T_{1}$ limited performance.  

These demonstrations highlight the potential for error-robust quantum control to have impact on the performance and design of quantum computers, and flow through to impact higher-level abstractions such as circuit compilation and quantum error correction~\cite{ball2020software}.  For instance, appropriately optimized pulses may permit relaxation of classical-crosstalk engineering tolerances, and permit full parallelization of all single-qubit unitaries as a default mode of operation without performance degradation.  Similarly, the error-rate-homogenization we have demonstrated here may simplify noise-aware compilation~\cite{vanmeter_2019, murali2019noiseadaptive, Tannu_2019} which nominally trades circuit complexity for improved gate errors.  

Since completing these experiments we have identified $\hat{\sigma}_{z}\hat{\sigma}_{z}$ couplings between certain qubit pairs which are not tabulated in backend properties and were accordingly not taken into account for our model-based approach to pulse optimization. We have also identified other minor discrepancies between IBM documentation for backend hardware parameters and observed values determined through system identification routines ({\em e.g.} sign and magnitude of tabulated qubit anharmonicities). In response, we have deployed autonomous reinforcement learning via Qiskit Pulse to discover optimized pulses without the need for complete models of the system Hamiltonian.  Initial results finding high-fidelity optimized pulses are exceptionally promising and will be the subject of a future publication.  Combining complete system identification and online optimization to compensate for unaccounted Hamiltonian terms provides a pathway to dramatically expand the performance of these solutions. Overall we hope that these demonstrations open new opportunities for complex low-level control strategies to augment the performance of near-term quantum computers.

\begin{acknowledgments}
We acknowledge the use of IBM Quantum services for this work. The views expressed are those of the authors, and do not reflect the official policy or position of IBM or the IBM Quantum team. The authors also acknowledge Nathan Earnest-Noble for technical discussions and his support on using Qiskit Pulse on IBM Quantum devices. 
\end{acknowledgments}

\bibliography{main.bbl}

\clearpage \newpage
\appendix
\begin{widetext}

\newpage
\section{Comparison between fitted and hardware-reported 
\texorpdfstring{$T_1$}{TEXT}
}
\label{app:T1}

At every daily hardware calibration, the decay time $T_1$ for each qubit in the backend is reported. These values are shown as gray circles in Fig.~\cref{fig:T1} for the whole period of our August experiments on {\sl Valencia}. For the IBM default pulse, the fitted decay times substantially underestimate the reported backend values across the device. By contrast, the fitted values extracted with robust pulses approximately agree with tabulated $T_1$ for most qubits and experiments (Fig.~\cref{fig:T1}, top row). Agreement gets worse with parallel implementation of pulses, indicating the existence of an extra decaying process.   

\begin{figure*}[htb]
\includegraphics[width=\textwidth]{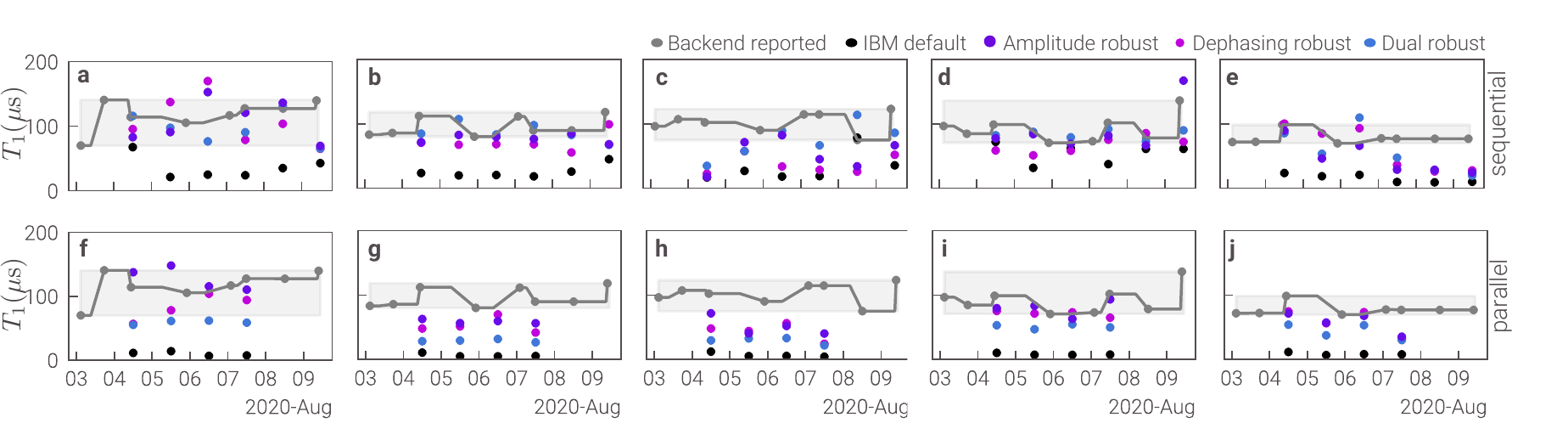}
\caption{Tabulated (gray circles) and fitted (colored circles) $T_1$ during the week of the experiments on {\sl Valencia} for qubits $0$ to $4$ (from left to right). The shaded area corresponds to the range between the minimum and maximum $T_1$ reported in that week. The fitted decay time is extracted from $T_1 = \tau_{g}/\beta$ using the gate durations and the tabulated $\beta$ or $\beta^{\parallel}$ values in~\cref{app:data_sets} for the serial (top row) and parallel (bottom row) experiments.}
\label{fig:T1}
\end{figure*}


\section{Optimized pulses implemented on {\sl Valencia}}\label{app:valencia_pulses}


The default high-fidelity gates on the IBM-Q hardware are obtained through a combination of DRAG pulses and virtual $Z$ rotations~\cite{McKay:2017}. In Qiskit, $R_{x}(\pi)$ and $R_{x}(\pi/2)$ pulses of this kind are called with the commands: 
\input{A60F5817ACA7E41A8C0C04C54841440E50EF3049E275D339AF879205FC52C1A0.pygtex}

We note that the IBM backend performs modifications of analytically defined DRAG pulses for each qubit in a manner that is undisclosed. Throughout this work ``Default'' pulses are called in this way without further calibration or modification.  Optimized pulses used throughout this work are described in detail in~\cref{fig:allpulses}.

\begin{figure*}[!]
\includegraphics[width=\textwidth]{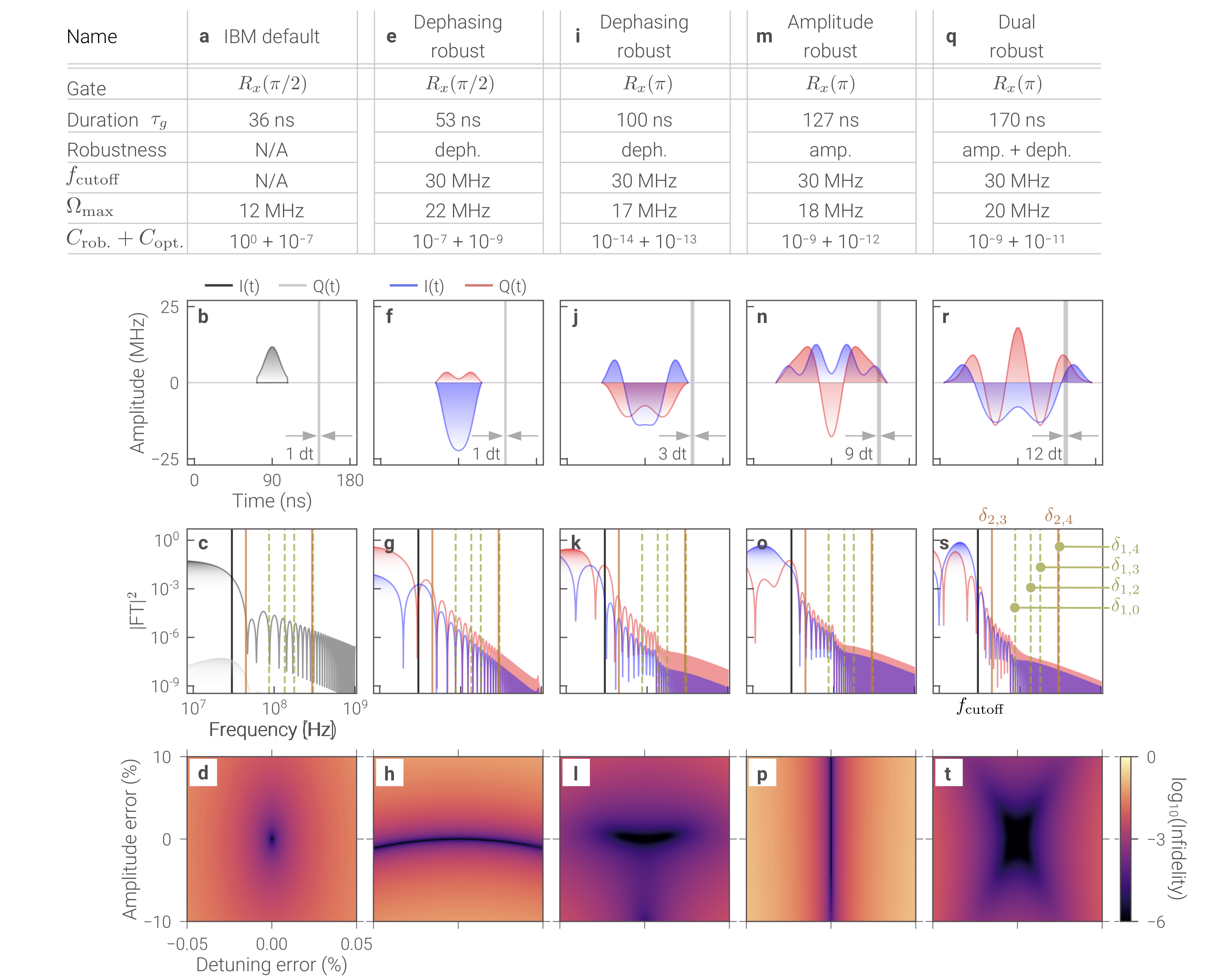}
\caption{Description and comparative analysis of pulse waveforms identified in the main text implemented on {\sl Valencia} backend. 
(a-d) IBM default $R_{x}(\frac{\pi}{2})$ pulse. 
(e-h) Q-CTRL dephasing-robust $R_{x}(\frac{\pi}{2})$ pulse. 
(i-l) Q-CTRL dephasing-robust $R_{x}(\pi)$ pulse. 
(m-p) Q-CTRL amplitude-robust $R_{x}(\pi)$ pulse. 
(q-t) Q-CTRL dual dephasing- and amplitude-robust $R_{x}(\pi)$ pulse. 
(a,e,i,m,q) Tabulated characteristics of IBM default DRAG compared with optimized Q-CTRL pulses. Q-CTRL pulses were optimized for robustness against dephasing/amplitutde errors (row 3) subject to band-limited (row 4) and bound-amplitude constraints (row 5). Here $f_\text{cutoff}$ refers to the 3 dB cutoff frequency associated with a low-pass sinc filter, and $\Omega_\text{max}$ is the maximum Rabi-rate over the pulse duration. Optimization costs are shown in row 6, with cost functions $C_\text{opt.}$  and $C_\text{rob.}$ defined in Ball {\em et. al}, 2020~\cite{ball2020software}. For comparison, panel (a) shows these metrics computed for the default pulse.
(b,f,j,n,r) IQ waveforms for IBM default DRAG pulse (grey) compared with optimized Q-CTRL pulses (red/blue). Durations for each pulse are tabulated in row 2 of the table above. Vertical gray shading indicates the segment duration for each pulse, where $dt = 0.22$ ns is the minimum sample time permitted by the IBM Valencia backend. 
(c,g,k,o,s) Fourier analysis of IQ waveforms for each pulse plotted above. Decay in pulse spectral features is consistent with low-pass sinc filter with $f_\text{cutoff} = 30$ MHz (vertical black lines). Spectra are overlaid with leakage transitions associated with detunings from higher-level states of neighbouring qubits~\cite{TheisPRA2016}, showing spectral overlap comparable with the default pulse. For qubit indexes $i\in\{0,1,2,3,4\}$ on Valencia, these transitions are defined as 
$\delta_{1,i}\equiv(\omega_{i}-\omega_{1})/2\pi$ 
and 
$\delta_{2,i}\equiv (2(\omega_{i}-\omega_{1})-\chi_{i})/2\pi$ where $\omega_{i}$ and $\chi_{i}$ are respectively the resonant frequency and anharmonicities for qubit $i$. 
(d,h,l,p,t) Log-scale density plots showing robustness properties for each pulse. Dephasing and amplitude robustness is indicated by dark regions in the 2D domain corresponding to low infidelities while scanning over these errors. Infidelities were calculated assuming a 2-level system model incorporating offset detuning- and amplitude-errors, plotted as percentages of the nominal qubit frequency ($\sim$ 5 GHz) and ideal per-segment IQ amplitudes respectively. Q-CTRL solutions suppress these errors by orders of magnitude compared to the default pulse. Dephasing- and amplitude-robustness is seen by the extended regions of low infidelity along these axes. The dual-robust solution plotted in panel (r) shows robustness in both directions. These are to be compared with the IBM default DRAG pulse in panel (d), showing sharp increase in infidelity with both these errors. 
}
\label{fig:allpulses}
\end{figure*}

\newpage
\section{Supplementary data sets}
\label{app:data_sets}

In this Appendix we provide tabulated data for the rotation error $\epsilon_{r}$, decay constant $\beta$, and total error $\epsilon$ extracted from experiments summarized in~\cref{fig:device_wide_performance_sequential} and~\cref{fig:device_wide_performance_parallel}.  For each day on which measurements were performed, and for each qubit, we additionally calculate average device performance and variance over both time and across devices. Data are tabulated for experiments using both serialized operations and parallelized operations subject to crosstalk.
\newpage

\setlength{\tabcolsep}{9pt}

\begin{table*}\centering
\begin{tabular}{@{}l@{\hspace{-1ex}}rcccccc@{}cc}
\cline{1-7} \cline{9-10}
 &&\multicolumn{5}{c}{\textbf{Qubit}} & & \multicolumn{2}{c}{\textbf{System-averaged data}}  \\
 \cline{3-7}
 & & 0 & 1 &2 &3 &4 &&$\boldsymbol{\langle \epsilon \rangle_q}$&$\boldsymbol {\sigma_q}$\\
\cline{1-7} \cline{9-10} \\
\multirow{3}{*}{\rotatebox[origin=c]{90}{Aug 4}} & \multicolumn{1}{c|}{$\epsilon_r$} & $5.47 \times 10^{-3}$ & $1.21 \times 10^{-2}$ & $2.96 \times 10^{-3}$ & $6.14 \times 10^{-3}$ & \multicolumn{1}{c|}{$5.78 \times 10^{-3}$} 
 & & &  \\
 &\multicolumn{1}{c|}{$\beta$} & $1.05 \times 10^{-3}$ & $3.03 \times 10^{-3}$ & $4.28 \times 10^{-3}$ & $9.88 \times 10^{-4}$ & \multicolumn{1}{c|}{$3.03 \times 10^{-3}$} & & & \\
&\multicolumn{1}{c|}{$\epsilon$} & $5.57 \times 10^{-3}$ & $1.25 \times 10^{-2}$ & $5.20 \times 10^{-3}$ & $6.22 \times 10^{-3}$ & \multicolumn{1}{c|}{$6.52 \times 10^{-3}$} & & $7.21 \times 10^{-3}$& $2.69 \times 10^{-3}$   \\
\\
\multirow{3}{*}{\rotatebox[origin=c]{90}{Aug 5}} &\multicolumn{1}{c|}{$\epsilon_r$} & $1.08 \times 10^{-2}$ & $4.89 \times 10^{-3}$ & $1.69 \times 10^{-2}$ & $5.84 \times 10^{-3}$ & \multicolumn{1}{c|}{$1.60 \times 10^{-2}$}
 &&& \\ 
&\multicolumn{1}{c|}{$\beta$} & $3.34 \times 10^{-3}$ & $3.58 \times 10^{-3}$ & $2.65 \times 10^{-3}$ & $2.25 \times 10^{-3}$ & \multicolumn{1}{c|}{$3.84 \times 10^{-3}$} &&&\\
&\multicolumn{1}{c|}{$\epsilon$} & $1.13 \times 10^{-2}$ & $6.06 \times 10^{-3}$ & $1.71 \times 10^{-2}$ & $6.25 \times 10^{-3}$ & \multicolumn{1}{c|}{$1.64 \times 10^{-2}$}  & & $1.14 \times 10^{-2}$& $4.76 \times 10^{-3}$ \\
\\
\multirow{3}{*}{\rotatebox[origin=c]{90}{Aug 6}} &\multicolumn{1}{c|}{$\epsilon_r$} & $8.36 \times 10^{-3}$ & $3.80 \times 10^{-3}$ & $7.71 \times 10^{-3}$ & $3.58 \times 10^{-3}$ & \multicolumn{1}{c|}{$1.37 \times 10^{-2}$} \\ 
&\multicolumn{1}{c|}{$\beta$} &$2.85 \times 10^{-3}$ & $3.48 \times 10^{-3}$ & $3.94 \times 10^{-3}$ & $1.15 \times 10^{-3}$ & \multicolumn{1}{c|}{$3.44 \times 10^{-3}$} \\
&\multicolumn{1}{c|}{$\epsilon$}& $8.83 \times 10^{-3}$ & $5.15 \times 10^{-3}$ & $8.66 \times 10^{-3}$ & $3.76 \times 10^{-3}$ & \multicolumn{1}{c|}{$1.41 \times 10^{-2}$} & & $8.10 \times 10^{-3}$& $3.60 \times 10^{-3}$ \\
\\
\multirow{3}{*}{\rotatebox[origin=c]{90}{Aug 7}} &\multicolumn{1}{c|}{$\epsilon_r$} & $7.62 \times 10^{-3}$ & $5.68 \times 10^{-3}$ & $8.71 \times 10^{-3}$ & $3.29 \times 10^{-3}$ & \multicolumn{1}{c|}{$1.83 \times 10^{-2}$} \\ 
&\multicolumn{1}{c|}{$\beta$}&$2.96 \times 10^{-3}$ & $3.88 \times 10^{-3}$ & $3.80 \times 10^{-3}$ & $1.91 \times 10^{-3}$ & \multicolumn{1}{c|}{$7.47 \times 10^{-3}$} \\
&\multicolumn{1}{c|}{$\epsilon$} &$8.18 \times 10^{-3}$ & $6.88 \times 10^{-3}$ & $9.50 \times 10^{-3}$ & $3.80 \times 10^{-3}$ & \multicolumn{1}{c|}{$1.98 \times 10^{-2}$}& & $9.63 \times 10^{-3}$& $5.42 \times 10^{-3}$ \\
\\
\multirow{3}{*}{\rotatebox[origin=c]{90}{Aug 8}} &\multicolumn{1}{c|}{$\epsilon_r$} & $3.36 \times 10^{-3}$ & $7.46 \times 10^{-3}$ & $3.87 \times 10^{-3}$ & $3.49 \times 10^{-3}$ & \multicolumn{1}{c|}{$1.04 \times 10^{-2}$} \\ 
&\multicolumn{1}{c|}{$\beta$}&$2.03 \times 10^{-3}$ & $2.77 \times 10^{-3}$ & $9.05 \times 10^{-4}$ & $1.16 \times 10^{-3}$ & \multicolumn{1}{c|}{$7.90 \times 10^{-3}$} \\
&\multicolumn{1}{c|}{$\epsilon$} &$3.93 \times 10^{-3}$ & $7.95 \times 10^{-3}$ & $3.98 \times 10^{-3}$ & $3.68 \times 10^{-3}$ & \multicolumn{1}{c|}{$1.31 \times 10^{-2}$} & & $6.52 \times 10^{-3}$& $3.64 \times 10^{-3}$ \\
\\
\multirow{3}{*}{\rotatebox[origin=c]{90}{Aug 9}} &\multicolumn{1}{c|}{$\epsilon_r$} & $3.41 \times 10^{-3}$ & $5.89 \times 10^{-3}$ & $2.16 \times 10^{-3}$ & $4.77 \times 10^{-3}$ & \multicolumn{1}{c|}{$6.62 \times 10^{-3}$} \\ 
&\multicolumn{1}{c|}{$\beta$}&$1.66 \times 10^{-3}$ & $1.58 \times 10^{-3}$ & $2.00 \times 10^{-3}$ & $1.16 \times 10^{-3}$ & \multicolumn{1}{c|}{$7.23 \times 10^{-3}$} \\
&\multicolumn{1}{c|}{$\epsilon$} &$3.79 \times 10^{-3}$ & $6.10 \times 10^{-3}$ & $2.94 \times 10^{-3}$ & $4.91 \times 10^{-3}$ & \multicolumn{1}{c|}{$9.80 \times 10^{-3}$}& & $5.51 \times 10^{-3}$& $2.40 \times 10^{-3}$ \\
\\
\cline{1-7} \cline{9-10}
\multirow{4}{*}{\rotatebox[origin=c]{90}{\parbox[r]{1.5 cm}{{\textbf{Time  averaged}}}}} &&&&&&&& \multicolumn{2}{|c|}{\textbf{Total average}} \\
& \multicolumn{1}{c|}{$\boldsymbol{\langle \epsilon \rangle_t}$} 
&$6.94 \times 10^{-3}$ & $7.44 \times 10^{-3}$ & $7.90 \times 10^{-3}$ & $4.77 \times 10^{-3}$ & \multicolumn{1}{c|}{$1.33 \times 10^{-2}$} & & \multicolumn{2}{|c|}{$\boldsymbol{\langle \epsilon \rangle}= 8.07\times 10^{-3}$}
\\
& \multicolumn{1}{c|}{$\boldsymbol {\sigma_t}$}& $2.74 \times 10^{-3}$ & $2.42 \times 10^{-3}$  & $4.76 \times 10^{-3}$ & $1.12 \times 10^{-3}$ & \multicolumn{1}{c|}{$4.30 \times 10^{-3}$} & &\multicolumn{2}{|c|}{$\boldsymbol{\langle \sigma_t \rangle_q}= 3.07\times 10^{-3}$} 
\\
&&&&&&&&\multicolumn{2}{|c|}{$\boldsymbol{\langle \sigma_q \rangle_t}= 3.75\times 10^{-3}$}
\\
\cline{1-7} \cline{9-10}
\end{tabular}
\caption{Rotation error ($\epsilon_r$), population-decay error per gate ($\beta$), and total effective error per gate ($\epsilon$) for all qubits and experimental runs from 4 to 9 August with serial application of the $R_x(\pi)$ gate using the IBM default pulse. The last two columns contain, respectively, the total error averaged over the qubits $\langle \epsilon \rangle_q$ and their deviations $\sigma_q$ for each day. The bottom two rows show the time-averaged errors $\langle \epsilon \rangle_t$ for each qubit and their deviations, $\sigma_t$, respectively. The total average, $\langle \epsilon \rangle=\langle \epsilon \rangle_{q,t}$, as well as the averages of the deviations, $\langle \sigma_t \rangle_q$ and $\langle \sigma_q \rangle_t$, are shown in the box at the bottom right of the table.
\label{table_1}}
\end{table*}

\begin{table*}\centering
\begin{tabular}{@{}l@{\hspace{-1ex}}rcccccc@{}cc}
\cline{1-7} \cline{9-10}
 &&\multicolumn{5}{c}{\textbf{Qubit}} & & \multicolumn{2}{c}{\textbf{System-averaged data}}  \\
 \cline{3-7}
 & & 0 & 1 &2 &3 &4 &&$\boldsymbol{\langle \epsilon \rangle_q}$&$\boldsymbol {\sigma_q}$\\
\cline{1-7} \cline{9-10} \\
\multirow{3}{*}{\rotatebox[origin=c]{90}{Aug 4}} & \multicolumn{1}{c|}{$\epsilon_r$} & $2.13 \times 10^{-3}$ & $2.58 \times 10^{-8}$ & $2.25 \times 10^{-8}$ & $2.78 \times 10^{-4}$ & \multicolumn{1}{c|}{$3.28 \times 10^{-3}$}
 & & &  \\
 &\multicolumn{1}{c|}{$\beta$} & $1.54 \times 10^{-3}$ & $1.79 \times 10^{-3}$ & $7.30 \times 10^{-3}$ & $1.65 \times 10^{-3}$ & \multicolumn{1}{c|}{$1.43 \times 10^{-3}$} & & & \\
&\multicolumn{1}{c|}{$\epsilon$} & $2.63 \times 10^{-3}$ & $1.79 \times 10^{-3}$ & $7.30 \times 10^{-3}$ & $1.67 \times 10^{-3}$ & \multicolumn{1}{c|}{$3.58 \times 10^{-3}$} & & $3.39 \times 10^{-3}$& $2.07 \times 10^{-3}$   \\
\\
\multirow{3}{*}{\rotatebox[origin=c]{90}{Aug 5}} &\multicolumn{1}{c|}{$\epsilon_r$} & $1.16 \times 10^{-7}$ & $7.11 \times 10^{-4}$ & $1.33 \times 10^{-3}$ & $2.14 \times 10^{-8}$ &\multicolumn{1}{c|}{$2.38 \times 10^{-3}$}
 &&& \\ 
&\multicolumn{1}{c|}{$\beta$} & $1.41 \times 10^{-3}$ & $1.55 \times 10^{-3}$ & $1.79 \times 10^{-3}$ & $1.52 \times 10^{-3}$ & \multicolumn{1}{c|}{$2.77 \times 10^{-3}$} &&&\\
&\multicolumn{1}{c|}{$\epsilon$} & $1.41 \times 10^{-3}$ & $1.70 \times 10^{-3}$ & $2.23 \times 10^{-3}$ & $1.52 \times 10^{-3}$ & \multicolumn{1}{c|}{$3.66 \times 10^{-3}$}  & & $2.10 \times 10^{-3}$& $8.25 \times 10^{-4}$ \\
\\
\multirow{3}{*}{\rotatebox[origin=c]{90}{Aug 6}} &\multicolumn{1}{c|}{$\epsilon_r$} & $1.57 \times 10^{-3}$ & $1.52 \times 10^{-8}$ & $1.52 \times 10^{-3}$ & $1.17 \times 10^{-8}$ & \multicolumn{1}{c|}{$2.55 \times 10^{-3}$} \\ 
&\multicolumn{1}{c|}{$\beta$} &$8.38 \times 10^{-4}$ & $1.61 \times 10^{-3}$ & $1.55 \times 10^{-3}$ & $1.86 \times 10^{-3}$ & \multicolumn{1}{c|}{$1.94 \times 10^{-3}$} \\
&\multicolumn{1}{c|}{$\epsilon$}& $1.78 \times 10^{-3}$ & $1.61 \times 10^{-3}$ & $2.18 \times 10^{-3}$ & $1.86 \times 10^{-3}$ & \multicolumn{1}{c|}{$3.21 \times 10^{-3}$} & & $2.13 \times 10^{-3}$& $5.71 \times 10^{-4}$ \\
\\
\multirow{3}{*}{\rotatebox[origin=c]{90}{Aug 7}} &\multicolumn{1}{c|}{$\epsilon_r$} & $4.63 \times 10^{-3}$ & $1.09 \times 10^{-8}$ & $6.58 \times 10^{-8}$ & $1.14 \times 10^{-3}$ & \multicolumn{1}{c|}{$1.47 \times 10^{-8}$} \\ 
&\multicolumn{1}{c|}{$\beta$}&$1.06 \times 10^{-3}$ & $1.67 \times 10^{-3}$ & $2.84 \times 10^{-3}$ & $1.57 \times 10^{-3}$ &  \multicolumn{1}{c|}{$4.45 \times 10^{-3}$} \\
&\multicolumn{1}{c|}{$\epsilon$} &$4.75 \times 10^{-3}$ & $1.67 \times 10^{-3}$ & $2.84 \times 10^{-3}$ & $1.94 \times 10^{-3}$ & \multicolumn{1}{c|}{$4.45 \times 10^{-3}$}& & $3.13 \times 10^{-3}$& $1.26 \times 10^{-3}$ \\
\\
\multirow{3}{*}{\rotatebox[origin=c]{90}{Aug 8}} &\multicolumn{1}{c|}{$\epsilon_r$} & $3.45 \times 10^{-3}$ & $7.26 \times 10^{-8}$ & $1.09 \times 10^{-7}$ & $4.40 \times 10^{-8}$ & \multicolumn{1}{c|}{$1.21 \times 10^{-7}$} \\ 
&\multicolumn{1}{c|}{$\beta$}&$9.39 \times 10^{-4}$ & $1.51 \times 10^{-3}$ & $3.77 \times 10^{-3}$ & $1.92 \times 10^{-3}$ & \multicolumn{1}{c|}{$4.50 \times 10^{-3}$} \\
&\multicolumn{1}{c|}{$\epsilon$} &$3.57 \times 10^{-3}$ & $1.51 \times 10^{-3}$ & $3.77 \times 10^{-3}$ & $1.92 \times 10^{-3}$ & \multicolumn{1}{c|}{$4.50 \times 10^{-3}$} & & $3.05 \times 10^{-3}$& $1.14 \times 10^{-3}$ \\
\\
\multirow{3}{*}{\rotatebox[origin=c]{90}{Aug 9}} &\multicolumn{1}{c|}{$\epsilon_r$} & $1.16 \times 10^{-3}$ & $9.28 \times 10^{-8}$ & $8.17 \times 10^{-9}$ & $2.04 \times 10^{-3}$ & \multicolumn{1}{c|}{$6.32 \times 10^{-8}$} \\ 
&\multicolumn{1}{c|}{$\beta$}&$1.84 \times 10^{-3}$ & $1.87 \times 10^{-3}$ & $1.92 \times 10^{-3}$ & $7.65 \times 10^{-4}$ & \multicolumn{1}{c|}{$5.30 \times 10^{-3}$} \\
&\multicolumn{1}{c|}{$\epsilon$} &$2.18 \times 10^{-3}$ & $1.87 \times 10^{-3}$ & $1.92 \times 10^{-3}$ & $2.18 \times 10^{-3}$ & \multicolumn{1}{c|}{$5.30 \times 10^{-3}$}& & $2.69 \times 10^{-3}$& $1.31 \times 10^{-3}$ \\
\\
\cline{1-7} \cline{9-10}
\multirow{4}{*}{\rotatebox[origin=c]{90}{\parbox[r]{1.5 cm}{{\textbf{Time  averaged}}}}} &&&&&&&&  \multicolumn{2}{|c|}{\textbf{Total averages}}\\
& \multicolumn{1}{c|}{$\boldsymbol{\langle \epsilon \rangle_t}$} 
&$2.72 \times 10^{-3}$ & $1.69 \times 10^{-3}$ & $3.37 \times 10^{-3}$ & $1.85 \times 10^{-3}$ & \multicolumn{1}{c|}{$4.11 \times 10^{-3}$} & & \multicolumn{2}{|c|}{$\boldsymbol{\langle \epsilon \rangle}= 2.75\times 10^{-3}$}
\\
& \multicolumn{1}{c|}{$\boldsymbol {\sigma_t}$}& $1.14 \times 10^{-3}$ & $1.18 \times 10^{-4}$  & $1.86 \times 10^{-3}$ & $2.07 \times 10^{-4}$ & \multicolumn{1}{c|}{$7.04 \times 10^{-4}$} & &\multicolumn{2}{|c|}{$\boldsymbol{\langle \sigma_t \rangle_q}= 8.05\times 10^{-4}$}  
\\
&&&&&&&&\multicolumn{2}{|c|}{$\boldsymbol{\langle \sigma_q \rangle_t}= 1.20\times 10^{-3}$} 
\\
\cline{1-7}\cline{9-10}
\end{tabular}
\caption{Same as in~\cref{table_1} but for the amplitude-robust pulse.}
\end{table*}

\begin{table*}\centering
\begin{tabular}{@{}l@{\hspace{-1ex}}rcccccc@{}cc}
\cline{1-7} \cline{9-10}
 &&\multicolumn{5}{c}{\textbf{Qubit}} & & \multicolumn{2}{c}{\textbf{System-averaged data}}  \\
 \cline{3-7}
 & & 0 & 1 &2 &3 &4 &&$\boldsymbol{\langle \epsilon \rangle_q}$&$\boldsymbol {\sigma_q}$\\
\cline{1-7} \cline{9-10} \\
\multirow{3}{*}{\rotatebox[origin=c]{90}{Aug 4}} & \multicolumn{1}{c|}{$\epsilon_r$} & $8.53 \times 10^{-8}$ & $3.16 \times 10^{-8}$ & $1.27 \times 10^{-10}$ & $8.30 \times 10^{-8}$& \multicolumn{1}{c|}{$7.40 \times 10^{-8}$} & & &  \\
 &\multicolumn{1}{c|}{$\beta$} & $1.04 \times 10^{-3}$ & $1.41 \times 10^{-3}$ & $4.43 \times 10^{-3}$ & $1.70 \times 10^{-3}$ & \multicolumn{1}{c|}{$9.92 \times 10^{-4}$} & & & \\
&\multicolumn{1}{c|}{$\epsilon$} & $1.04 \times 10^{-3}$ & $1.41 \times 10^{-3}$ & $4.43 \times 10^{-3}$ & $1.70 \times 10^{-3}$ & \multicolumn{1}{c|}{$9.92 \times 10^{-4}$} & & $1.92 \times 10^{-3}$& $1.29 \times 10^{-3}$   \\
\\
\multirow{3}{*}{\rotatebox[origin=c]{90}{Aug 5}} &\multicolumn{1}{c|}{$\epsilon_r$} & $1.92 \times 10^{-3}$ & $3.47 \times 10^{-3}$ & $1.60 \times 10^{-3}$ & $2.07 \times 10^{-3}$ & \multicolumn{1}{c|}{$2.49 \times 10^{-3}$} &&& \\ 
&\multicolumn{1}{c|}{$\beta$} & $7.25 \times 10^{-4}$ & $1.47 \times 10^{-3}$ & $1.73 \times 10^{-3}$ & $1.95 \times 10^{-3}$ & \multicolumn{1}{c|}{$1.17 \times 10^{-3}$} &&&\\
&\multicolumn{1}{c|}{$\epsilon$} & $2.06 \times 10^{-3}$ & $3.77 \times 10^{-3}$ & $2.36 \times 10^{-3}$ & $2.84 \times 10^{-3}$ & \multicolumn{1}{c|}{$2.75 \times 10^{-3}$}  & & $2.75 \times 10^{-3}$& $5.81 \times 10^{-4}$ \\
\\
\multirow{3}{*}{\rotatebox[origin=c]{90}{Aug 6}} &\multicolumn{1}{c|}{$\epsilon_r$} & $1.93 \times 10^{-3}$ & $5.08 \times 10^{-3}$ & $2.44 \times 10^{-8}$ & $3.04 \times 10^{-9}$ & \multicolumn{1}{c|}{$7.51 \times 10^{-4}$} \\ 
&\multicolumn{1}{c|}{$\beta$} &$5.86 \times 10^{-4}$ & $1.46 \times 10^{-3}$ & $2.96 \times 10^{-3}$ & $1.71 \times 10^{-3}$ & \multicolumn{1}{c|}{$1.07 \times 10^{-3}$} \\
&\multicolumn{1}{c|}{$\epsilon$}& $2.02 \times 10^{-3}$ & $5.29 \times 10^{-3}$ & $2.96 \times 10^{-3}$ & $1.71 \times 10^{-3}$ & \multicolumn{1}{c|}{$1.31 \times 10^{-3}$} & & $2.66 \times 10^{-3}$& $1.42 \times 10^{-3}$ \\
\\
\multirow{3}{*}{\rotatebox[origin=c]{90}{Aug 7}} &\multicolumn{1}{c|}{$\epsilon_r$} & $9.60 \times 10^{-4}$ & $5.05 \times 10^{-3}$ & $4.79 \times 10^{-8}$ & $3.81 \times 10^{-8}$ & \multicolumn{1}{c|}{$1.26 \times 10^{-8}$} \\ 
&\multicolumn{1}{c|}{$\beta$}&$1.26 \times 10^{-3}$ & $1.46 \times 10^{-3}$ & $3.49 \times 10^{-3}$ & $1.32 \times 10^{-3}$ & \multicolumn{1}{c|}{$2.76 \times 10^{-3}$} \\
&\multicolumn{1}{c|}{$\epsilon$} &$1.59 \times 10^{-3}$ & $5.26 \times 10^{-3}$ & $3.49 \times 10^{-3}$ & $1.32 \times 10^{-3}$ & \multicolumn{1}{c|}{$2.76 \times 10^{-3}$}& & $2.88 \times 10^{-3}$& $1.42 \times 10^{-3}$ \\
\\
\multirow{3}{*}{\rotatebox[origin=c]{90}{Aug 8}} &\multicolumn{1}{c|}{$\epsilon_r$} & $1.25 \times 10^{-3}$ & $4.44 \times 10^{-3}$ & $3.71 \times 10^{-7}$ & $2.81 \times 10^{-8}$ & \multicolumn{1}{c|}{$2.55 \times 10^{-3}$} \\ 
&\multicolumn{1}{c|}{$\beta$}&$9.60 \times 10^{-4}$ & $1.79 \times 10^{-3}$ & $3.88 \times 10^{-3}$ & $1.16 \times 10^{-3}$ & \multicolumn{1}{c|}{$3.83 \times 10^{-3}$} \\
&\multicolumn{1}{c|}{$\epsilon$} &$1.58 \times 10^{-3}$ & $4.79 \times 10^{-3}$ & $3.88 \times 10^{-3}$ & $1.16 \times 10^{-3}$ & \multicolumn{1}{c|}{$4.60 \times 10^{-3}$} & & $3.20 \times 10^{-3}$& $1.53 \times 10^{-3}$ \\
\\
\multirow{3}{*}{\rotatebox[origin=c]{90}{Aug 9}} &\multicolumn{1}{c|}{$\epsilon_r$} & $2.63 \times 10^{-3}$ & $4.89 \times 10^{-3}$ & $7.83 \times 10^{-4}$ & $6.03 \times 10^{-4}$ & \multicolumn{1}{c|}{$5.80 \times 10^{-8}$} \\ 
&\multicolumn{1}{c|}{$\beta$}&$1.53 \times 10^{-3}$ & $1.00 \times 10^{-3}$ & $1.92 \times 10^{-3}$ & $1.38 \times 10^{-3}$ & \multicolumn{1}{c|}{$3.61 \times 10^{-3}$} \\
&\multicolumn{1}{c|}{$\epsilon$} &$3.04 \times 10^{-3}$ & $4.99 \times 10^{-3}$ & $2.07 \times 10^{-3}$ & $1.51 \times 10^{-3}$ & \multicolumn{1}{c|}{$3.61 \times 10^{-3}$}& & $3.04 \times 10^{-3}$& $1.22 \times 10^{-3}$ \\
\\
\cline{1-7} \cline{9-10}
\multirow{4}{*}{\rotatebox[origin=c]{90}{\parbox[r]{1.5 cm}{{\textbf{Time  averaged}}}}} &&&&&&&& \multicolumn{2}{|c|}{\textbf{Total average}} \\
& \multicolumn{1}{c|}{$\boldsymbol{\langle \epsilon \rangle_t}$} &$1.89 \times 10^{-3}$ & $4.25 \times 10^{-3}$ & $3.20 \times 10^{-3}$ & $1.71 \times 10^{-3}$ & \multicolumn{1}{c|}{$2.67 \times 10^{-3}$} & & \multicolumn{2}{|c|}{$\boldsymbol{\langle \epsilon \rangle}= 2.74\times 10^{-3}$}
\\
& \multicolumn{1}{c|}{$\boldsymbol {\sigma_t}$}& $6.16 \times 10^{-4}$ & $1.37 \times 10^{-3}$  & $8.27 \times 10^{-4}$ & $5.44 \times 10^{-4}$ & \multicolumn{1}{c|}{$1.25 \times 10^{-3}$} & &\multicolumn{2}{|c|}{$\boldsymbol{\langle \sigma_t \rangle_q}= 9.20\times 10^{-4}$}  
\\
&&&&&&&&\multicolumn{2}{|c|}{$\boldsymbol{\langle \sigma_q \rangle_t}= 1.24\times 10^{-3}$}
\\
\cline{1-7} \cline{9-10}
\end{tabular}
\caption{Same as in~\cref{table_1} but for the dephasing-robust pulse.}
\end{table*}

\begin{table*}\centering
\begin{tabular}{@{}l@{\hspace{-1ex}}rcccccc@{}cc}
\cline{1-7} \cline{9-10}
 &&\multicolumn{5}{c}{\textbf{Qubit}} & & \multicolumn{2}{c}{\textbf{System-averaged data}}  \\
 \cline{3-7}
 & & 0 & 1 &2 &3 &4 &&$\boldsymbol{\langle \epsilon \rangle_q}$&$\boldsymbol {\sigma_q}$\\
\cline{1-7} \cline{9-10} \\
\multirow{3}{*}{\rotatebox[origin=c]{90}{Aug 4}} & \multicolumn{1}{c|}{$\epsilon_r$} & $2.07 \times 10^{-8}$ & $6.62 \times 10^{-8}$ & $5.62 \times 10^{-8}$ & $4.81 \times 10^{-8}$ & \multicolumn{1}{c|}{$3.61 \times 10^{-8}$} & & &  \\
 &\multicolumn{1}{c|}{$\beta$} & $1.47 \times 10^{-3}$ & $2.01 \times 10^{-3}$ & $4.92 \times 10^{-3}$ & $2.08 \times 10^{-3}$ & \multicolumn{1}{c|}{$2.00 \times 10^{-3}$} & & & \\
&\multicolumn{1}{c|}{$\epsilon$} &$1.47 \times 10^{-3}$ & $2.01 \times 10^{-3}$ & $4.92 \times 10^{-3}$ & $2.08 \times 10^{-3}$ & \multicolumn{1}{c|}{$2.00 \times 10^{-3}$} & & $2.50 \times 10^{-3}$& $1.23 \times 10^{-3}$   \\
\\
\multirow{3}{*}{\rotatebox[origin=c]{90}{Aug 5}} &\multicolumn{1}{c|}{$\epsilon_r$} & $5.13 \times 10^{-8}$ & $1.08 \times 10^{-3}$ & $2.25 \times 10^{-3}$ & $5.13 \times 10^{-8}$ & \multicolumn{1}{c|}{$1.16 \times 10^{-7}$} &&& \\ 
&\multicolumn{1}{c|}{$\beta$} & $1.75 \times 10^{-3}$ & $1.59 \times 10^{-3}$ & $2.97 \times 10^{-3}$ & $1.95 \times 10^{-3}$ & \multicolumn{1}{c|}{$3.19 \times 10^{-3}$} &&&\\
&\multicolumn{1}{c|}{$\epsilon$} & $1.75 \times 10^{-3}$ & $1.92 \times 10^{-3}$ & $3.73 \times 10^{-3}$ & $1.95 \times 10^{-3}$ & \multicolumn{1}{c|}{$3.19 \times 10^{-3}$}  & & $2.51 \times 10^{-3}$& $7.98 \times 10^{-4}$ \\
\\
\multirow{3}{*}{\rotatebox[origin=c]{90}{Aug 6}} &\multicolumn{1}{c|}{$\epsilon_r$} & $1.08 \times 10^{-3}$ & $1.33 \times 10^{-3}$ & $9.33 \times 10^{-4}$ & $2.75 \times 10^{-8}$ & \multicolumn{1}{c|}{$3.11 \times 10^{-8}$} \\ 
&\multicolumn{1}{c|}{$\beta$} &$2.23 \times 10^{-3}$ & $2.04 \times 10^{-3}$ & $1.91 \times 10^{-3}$ & $2.16 \times 10^{-3}$ & \multicolumn{1}{c|}{$1.56 \times 10^{-3}$} \\
&\multicolumn{1}{c|}{$\epsilon$}& $2.48 \times 10^{-3}$ & $2.44 \times 10^{-3}$ & $2.13 \times 10^{-3}$ & $2.16 \times 10^{-3}$ & \multicolumn{1}{c|}{$1.56 \times 10^{-3}$} & & $2.15 \times 10^{-3}$& $3.31 \times 10^{-4}$ \\
\\
\multirow{3}{*}{\rotatebox[origin=c]{90}{Aug 7}} &\multicolumn{1}{c|}{$\epsilon_r$} & $1.44 \times 10^{-3}$ & $1.06 \times 10^{-3}$ & $3.49 \times 10^{-8}$ & $2.36 \times 10^{-3}$ & \multicolumn{1}{c|}{$7.29 \times 10^{-9}$} \\ 
&\multicolumn{1}{c|}{$\beta$}&$1.89 \times 10^{-3}$ & $1.74 \times 10^{-3}$ & $2.56 \times 10^{-3}$ & $1.86 \times 10^{-3}$ & \multicolumn{1}{c|}{$3.60 \times 10^{-3}$} \\
&\multicolumn{1}{c|}{$\epsilon$} &$2.37 \times 10^{-3}$ & $2.04 \times 10^{-3}$ & $2.56 \times 10^{-3}$ & $3.01 \times 10^{-3}$ & \multicolumn{1}{c|}{$3.60 \times 10^{-3}$}& & $2.71 \times 10^{-3}$& $5.42 \times 10^{-4}$ \\
\\
\multirow{3}{*}{\rotatebox[origin=c]{90}{Aug 8}} &\multicolumn{1}{c|}{$\epsilon_r$} & $3.92 \times 10^{-8}$ & $5.64 \times 10^{-8}$ & $3.30 \times 10^{-3}$ & $3.97 \times 10^{-9}$ & \multicolumn{1}{c|}{$5.07 \times 10^{-8}$} \\ 
&\multicolumn{1}{c|}{$\beta$}&$1.30 \times 10^{-3}$ & $2.06 \times 10^{-3}$ & $1.50 \times 10^{-3}$ & $2.35 \times 10^{-3}$ & \multicolumn{1}{c|}{$6.04 \times 10^{-3}$} \\
&\multicolumn{1}{c|}{$\epsilon$} &$1.30 \times 10^{-3}$ & $2.06 \times 10^{-3}$ & $3.62 \times 10^{-3}$ & $2.35 \times 10^{-3}$ & \multicolumn{1}{c|}{$6.04 \times 10^{-3}$} & & $3.07 \times 10^{-3}$& $1.66 \times 10^{-3}$ \\
\\
\multirow{3}{*}{\rotatebox[origin=c]{90}{Aug 9}} &\multicolumn{1}{c|}{$\epsilon_r$} & $1.57 \times 10^{-8}$ & $2.30 \times 10^{-8}$ & $2.07 \times 10^{-3}$ & $1.21 \times 10^{-3}$ & \multicolumn{1}{c|}{$1.96 \times 10^{-7}$} \\ 
&\multicolumn{1}{c|}{$\beta$}&$2.60 \times 10^{-3}$ & $2.53 \times 10^{-3}$ & $1.98 \times 10^{-3}$ & $1.90 \times 10^{-3}$ & \multicolumn{1}{c|}{$8.74 \times 10^{-3}$} \\
&\multicolumn{1}{c|}{$\epsilon$} &$2.60 \times 10^{-3}$ & $2.53 \times 10^{-3}$ & $2.86 \times 10^{-3}$ & $2.25 \times 10^{-3}$ & \multicolumn{1}{c|}{$8.74 \times 10^{-3}$}& & $3.70 \times 10^{-3}$& $2.27 \times 10^{-3}$ \\
\\
\cline{1-7} \cline{9-10}
\multirow{4}{*}{\rotatebox[origin=c]{90}{\parbox[r]{1.5 cm}{{\textbf{Time  averaged}}}}} &&&&&&&& \multicolumn{2}{|c|}{\textbf{Total average}} \\
& \multicolumn{1}{c|}{$\boldsymbol{\langle \epsilon \rangle_t}$} &$2.00 \times 10^{-3}$ & $2.17 \times 10^{-3}$ & $3.30 \times 10^{-3}$ & $2.30 \times 10^{-3}$ & \multicolumn{1}{c|}{$4.10 \times 10^{-3}$} & & \multicolumn{2}{|c|}{$\boldsymbol{\langle \epsilon \rangle}= 2.77\times 10^{-3}$}
\\
& \multicolumn{1}{c|}{$\boldsymbol {\sigma_t}$}& $5.10 \times 10^{-4}$ & $2.31 \times 10^{-4}$  & $9.15 \times 10^{-4}$ & $3.40 \times 10^{-4}$ & \multicolumn{1}{c|}{$2.34 \times 10^{-3}$} & &\multicolumn{2}{|c|}{$\boldsymbol{\langle \sigma_t \rangle_q}= 8.67\times 10^{-4}$}  
\\
&&&&&&&&\multicolumn{2}{|c|}{$\boldsymbol{\langle \sigma_q \rangle_t}= 1.14\times 10^{-3}$}
\\
\cline{1-7} \cline{9-10}
\end{tabular}
\caption{Same as in~\cref{table_1} but for the dual robust pulse.}
\end{table*}


\begin{table*}\centering
\begin{tabular}{@{}l@{\hspace{-1ex}}rcccccc@{}cc}
\cline{1-7} \cline{9-10}
 &&\multicolumn{5}{c}{\textbf{Qubit}} & & \multicolumn{2}{c}{\textbf{System-averaged data}}  \\
 \cline{3-7}
 & & 0 & 1 &2 &3 &4 &&$\boldsymbol{\langle \epsilon \rangle_q}$&$\boldsymbol {\sigma_q}$\\
\cline{1-7} \cline{9-10} \\
\multirow{3}{*}{\rotatebox[origin=c]{90}{Aug 4}} & \multicolumn{1}{c|}{$\epsilon_r$} &
$1.92 \times 10^{-3}$ & $2.04 \times 10^{-3}$ & $1.79 \times 10^{-3}$ & $2.24 \times 10^{-3}$ & \multicolumn{1}{c|}{$1.79 \times 10^{-3}$} & & &  \\
 &\multicolumn{1}{c|}{$\beta$} & $6.37 \times 10^{-3}$ & $6.69 \times 10^{-3}$ & $5.87 \times 10^{-3}$ & $6.99 \times 10^{-3}$ & \multicolumn{1}{c|}{$6.17 \times 10^{-3}$} & & & \\
&\multicolumn{1}{c|}{$\epsilon$} &$1.26 \times 10^{-2}$ & $1.64 \times 10^{-2}$ & $1.59 \times 10^{-2}$ & $1.16 \times 10^{-2}$ &    \multicolumn{1}{c|}{$1.43 \times 10^{-2}$} & & $1.42 \times 10^{-2}$& $1.86 \times 10^{-3}$   \\
\\
\multirow{3}{*}{\rotatebox[origin=c]{90}{Aug 5}} &\multicolumn{1}{c|}{$\epsilon_r$} & $2.50 \times 10^{-3}$ & $2.46 \times 10^{-3}$ & $3.16 \times 10^{-3}$ & $3.04 \times 10^{-3}$ & \multicolumn{1}{c|}{$2.24 \times 10^{-3}$} &&& \\ 
&\multicolumn{1}{c|}{$\beta$} & $5.13 \times 10^{-3}$ & $1.46 \times 10^{-2}$ & $1.39 \times 10^{-2}$ & $9.91 \times 10^{-3}$ &    \multicolumn{1}{c|}{$1.06 \times 10^{-2}$} &&&\\
&\multicolumn{1}{c|}{$\epsilon$} & $1.39 \times 10^{-2}$ & $1.46 \times 10^{-2}$ & $1.39 \times 10^{-2}$ & $9.91 \times 10^{-3}$ &    \multicolumn{1}{c|}{$1.80 \times 10^{-2}$}  & & $1.41 \times 10^{-2}$& $2.57 \times 10^{-3}$ \\
\\
\multirow{3}{*}{\rotatebox[origin=c]{90}{Aug 6}} &\multicolumn{1}{c|}{$\epsilon_r$} & $1.86 \times 10^{-3}$ & $2.15 \times 10^{-3}$ & $2.46 \times 10^{-3}$ & $2.23 \times 10^{-3}$ & \multicolumn{1}{c|}{$1.89 \times 10^{-3}$} \\ 
&\multicolumn{1}{c|}{$\beta$} &$1.09 \times 10^{-2}$ & $1.48 \times 10^{-2}$ & $1.32 \times 10^{-2}$ & $1.01 \times 10^{-2}$ &    \multicolumn{1}{c|}{$8.76 \times 10^{-3}$} \\
&\multicolumn{1}{c|}{$\epsilon$}& $1.09 \times 10^{-2}$ & $1.48 \times 10^{-2}$ & $1.32 \times 10^{-2}$ & $1.01 \times 10^{-2}$ &    \multicolumn{1}{c|}{$1.80 \times 10^{-2}$} & & $1.36 \times 10^{-2}$& $2.67 \times 10^{-3}$ \\
\\
\multirow{3}{*}{\rotatebox[origin=c]{90}{Aug 7}} &\multicolumn{1}{c|}{$\epsilon_r$} &$1.66 \times 10^{-3}$ & $2.68 \times 10^{-3}$ & $3.18 \times 10^{-3}$ & $4.03 \times 10^{-3}$ &    \multicolumn{1}{c|}{$3.62 \times 10^{-3}$} \\ 
&\multicolumn{1}{c|}{$\beta$}& $9.80 \times 10^{-3}$ & $1.35 \times 10^{-2}$ & $1.67 \times 10^{-2}$ & $9.47 \times 10^{-3}$ &  \multicolumn{1}{c|}{$9.14 \times 10^{-3}$} \\
&\multicolumn{1}{c|}{$\epsilon$} &$9.80 \times 10^{-3}$ & $1.35 \times 10^{-2}$ & $1.86 \times 10^{-2}$ & $9.47 \times 10^{-3}$ &    \multicolumn{1}{c|}{$1.71 \times 10^{-2}$}& & $1.37 \times 10^{-2}$& $3.70 \times 10^{-3}$ \\
\\
\cline{1-7} \cline{9-10}
\multirow{4}{*}{\rotatebox[origin=c]{90}{\parbox[r]{1.5 cm}{{\textbf{Time  averaged}}}}} &&&&&&&& \multicolumn{2}{|c|}{\textbf{Total average}} \\
& \multicolumn{1}{c|}{$\boldsymbol{\langle \epsilon \rangle_t}$}
&$1.21 \times 10^{-2}$ & $1.48 \times 10^{-2}$  & $1.54 \times 10^{-2}$ & $1.03 \times 10^{-2}$ & \multicolumn{1}{c|}{$1.68 \times 10^{-2}$} & & \multicolumn{2}{|c|}{$\boldsymbol{\langle \epsilon \rangle}= 1.39\times 10^{-2}$}
\\
& \multicolumn{1}{c|}{$\boldsymbol {\sigma_t}$}& $1.50 \times 10^{-3}$ & $1.07 \times 10^{-3}$ & $2.07 \times 10^{-3}$ & $8.06 \times 10^{-4}$ & \multicolumn{1}{c|}{$1.50 \times 10^{-3}$}
 & &\multicolumn{2}{|c|}{$\boldsymbol{\langle \sigma_t \rangle_q}= 1.39\times 10^{-3}$}  
\\
&&&&&&&&\multicolumn{2}{|c|}{$\boldsymbol{\langle \sigma_q \rangle_t}= 2.70\times 10^{-3}$}
\\
\cline{1-7} \cline{9-10}
\end{tabular}
\caption{Same as in~\cref{table_1} but for the case of parallel gate implementation. All fittings used~\cref{eq:Gausscosdecay} except for qubit 2 on 7 August where~\cref{eq:expcosdecay} provided a better fit.}
\end{table*}

\begin{table*}\centering
\begin{tabular}{@{}l@{\hspace{-1ex}}rcccccc@{}cc}
\cline{1-7} \cline{9-10}
 &&\multicolumn{5}{c}{\textbf{Qubit}} & & \multicolumn{2}{c}{\textbf{System-averaged data}}  \\
 \cline{3-7}
 & & 0 & 1 &2 &3 &4 &&$\boldsymbol{\langle \epsilon \rangle_q}$&$\boldsymbol {\sigma_q}$\\
\cline{1-7} \cline{9-10} \\
\multirow{3}{*}{\rotatebox[origin=c]{90}{Aug 4}} & \multicolumn{1}{c|}{$\epsilon_r$} &
$1.68 \times 10^{-3}$ & $4.04 \times 10^{-9}$ & $7.36 \times 10^{-8}$ & $1.55 \times 10^{-3}$ & \multicolumn{1}{c|}{$9.43 \times 10^{-8}$} & & &  \\
 &\multicolumn{1}{c|}{$\beta$} & $9.31 \times 10^{-4}$ & $2.04 \times 10^{-3}$ & $1.79 \times 10^{-3}$ & $1.61 \times 10^{-3}$ & \multicolumn{1}{c|}{$1.79 \times 10^{-3}$} & & & \\
&\multicolumn{1}{c|}{$\epsilon$} &$1.92 \times 10^{-3}$ & $2.04 \times 10^{-3}$ & $1.79 \times 10^{-3}$ & $2.24 \times 10^{-3}$ &   \multicolumn{1}{c|}{$1.79 \times 10^{-3}$} & & $1.95 \times 10^{-3}$& $1.69 \times 10^{-4}$   \\
\\
\multirow{3}{*}{\rotatebox[origin=c]{90}{Aug 5}} &\multicolumn{1}{c|}{$\epsilon_r$} & $2.35 \times 10^{-3}$ & $9.54 \times 10^{-4}$ & $7.52 \times 10^{-9}$ & $2.63 \times 10^{-3}$ & \multicolumn{1}{c|}{$1.62 \times 10^{-9}$} &&& \\ 
&\multicolumn{1}{c|}{$\beta$} & $8.65 \times 10^{-4}$ & $2.27 \times 10^{-3}$ & $3.16 \times 10^{-3}$ & $1.54 \times 10^{-3}$ &   \multicolumn{1}{c|}{$2.24 \times 10^{-3}$} &&&\\
&\multicolumn{1}{c|}{$\epsilon$} & $2.50 \times 10^{-3}$ & $2.46 \times 10^{-3}$ & $3.16 \times 10^{-3}$ & $3.04 \times 10^{-3}$ &   \multicolumn{1}{c|}{$2.24 \times 10^{-3}$}  & & $2.68 \times 10^{-3}$& $3.57 \times 10^{-4}$ \\
\\
\multirow{3}{*}{\rotatebox[origin=c]{90}{Aug 6}} &\multicolumn{1}{c|}{$\epsilon_r$} & $1.49 \times 10^{-3}$ & $1.58 \times 10^{-7}$ & $7.48 \times 10^{-8}$ & $9.33 \times 10^{-4}$ & \multicolumn{1}{c|}{$1.97 \times 10^{-8}$} \\ 
&\multicolumn{1}{c|}{$\beta$} &$1.11 \times 10^{-3}$ & $2.15 \times 10^{-3}$ & $2.46 \times 10^{-3}$ & $2.03 \times 10^{-3}$ &   \multicolumn{1}{c|}{$1.89 \times 10^{-3}$} \\
&\multicolumn{1}{c|}{$\epsilon$}& $1.86 \times 10^{-3}$ & $2.15 \times 10^{-3}$ & $2.46 \times 10^{-3}$ & $2.23 \times 10^{-3}$ &   \multicolumn{1}{c|}{$1.89 \times 10^{-3}$} & & $2.12 \times 10^{-3}$& $2.25 \times 10^{-4}$ \\
\\
\multirow{3}{*}{\rotatebox[origin=c]{90}{Aug 7}} &\multicolumn{1}{c|}{$\epsilon_r$} &$1.18 \times 10^{-3}$ & $1.41 \times 10^{-3}$ & $1.02 \times 10^{-7}$ & $3.79 \times 10^{-3}$ &   \multicolumn{1}{c|}{$1.72 \times 10^{-7}$} \\ 
&\multicolumn{1}{c|}{$\beta$}& $1.16 \times 10^{-3}$ & $2.28 \times 10^{-3}$ & $3.18 \times 10^{-3}$ & $1.37 \times 10^{-3}$ & \multicolumn{1}{c|}{$3.62 \times 10^{-3}$} \\
&\multicolumn{1}{c|}{$\epsilon$} &$1.66 \times 10^{-3}$ & $2.68 \times 10^{-3}$ & $3.18 \times 10^{-3}$ & $4.03 \times 10^{-3}$ &   \multicolumn{1}{c|}{$3.62 \times 10^{-3}$}& & $3.03 \times 10^{-3}$& $8.23 \times 10^{-4}$ \\
\\
\cline{1-7} \cline{9-10}
\multirow{4}{*}{\rotatebox[origin=c]{90}{\parbox[r]{1.5 cm}{{\textbf{Time  averaged}}}}} &&&&&&&& \multicolumn{2}{|c|}{\textbf{Total average}} \\
& \multicolumn{1}{c|}{$\boldsymbol{\langle \epsilon \rangle_t}$}
&$1.99 \times 10^{-3}$ & $2.33 \times 10^{-3}$  & $2.65 \times 10^{-3}$ & $2.89 \times 10^{-3}$ & \multicolumn{1}{c|}{$2.39 \times 10^{-3}$} & & \multicolumn{2}{|c|}{$\boldsymbol{\langle \epsilon \rangle}= 2.45\times 10^{-3}$}
\\
& \multicolumn{1}{c|}{$\boldsymbol {\sigma_t}$}& $3.15 \times 10^{-4}$ & $2.53 \times 10^{-4}$ & $5.75 \times 10^{-4}$ & $7.40 \times 10^{-4}$ & \multicolumn{1}{c|}{$7.34 \times 10^{-4}$}
 & &\multicolumn{2}{|c|}{$\boldsymbol{\langle \sigma_t \rangle_q}= 5.23\times 10^{-4}$}  
\\
&&&&&&&&\multicolumn{2}{|c|}{$\boldsymbol{\langle \sigma_q \rangle_t}= 3.93\times 10^{-4}$}
\\
\cline{1-7} \cline{9-10}
\end{tabular}
\caption{Same as in~\cref{table_1} but for the case of parallel gate implementation of the amplitude-robust pulse.}
\end{table*}

\begin{table*}\centering
\begin{tabular}{@{}l@{\hspace{-1ex}}rcccccc@{}cc}
\cline{1-7} \cline{9-10}
 &&\multicolumn{5}{c}{\textbf{Qubit}} & & \multicolumn{2}{c}{\textbf{System-averaged data}}  \\
 \cline{3-7}
 & & 0 & 1 &2 &3 &4 &&$\boldsymbol{\langle \epsilon \rangle_q}$&$\boldsymbol {\sigma_q}$\\
\cline{1-7} \cline{9-10} \\
\multirow{3}{*}{\rotatebox[origin=c]{90}{Aug 4}} & \multicolumn{1}{c|}{$\epsilon_r$} &
$2.68 \times 10^{-3}$ & $2.72 \times 10^{-3}$ & $2.43 \times 10^{-3}$ & $3.03 \times 10^{-3}$ & \multicolumn{1}{c|}{$3.20 \times 10^{-3}$} & & &  \\
 &\multicolumn{1}{c|}{$\beta$} & $1.76 \times 10^{-3}$ & $2.07 \times 10^{-3}$ & $2.07 \times 10^{-3}$ & $1.32 \times 10^{-3}$ &\multicolumn{1}{c|}{$1.34 \times 10^{-3}$} & & & \\
&\multicolumn{1}{c|}{$\epsilon$} &$3.20 \times 10^{-3}$ & $3.42 \times 10^{-3}$ & $3.20 \times 10^{-3}$ & $3.30 \times 10^{-3}$ &  \multicolumn{1}{c|}{$3.46 \times 10^{-3}$} & & $3.32 \times 10^{-3}$& $1.09 \times 10^{-4}$   \\
\\
\multirow{3}{*}{\rotatebox[origin=c]{90}{Aug 5}} &\multicolumn{1}{c|}{$\epsilon_r$} & $3.04 \times 10^{-3}$ & $2.76 \times 10^{-3}$ & $2.47 \times 10^{-3}$ & $3.13 \times 10^{-3}$ & \multicolumn{1}{c|}{$2.43 \times 10^{-3}$} &&& \\ 
&\multicolumn{1}{c|}{$\beta$} & $1.28 \times 10^{-3}$ & $1.94 \times 10^{-3}$ & $2.26 \times 10^{-3}$ & $1.40 \times 10^{-3}$ &  \multicolumn{1}{c|}{$1.77 \times 10^{-3}$} &&&\\
&\multicolumn{1}{c|}{$\epsilon$} & $3.30 \times 10^{-3}$ & $3.38 \times 10^{-3}$ & $3.34 \times 10^{-3}$ & $3.43 \times 10^{-3}$ &  \multicolumn{1}{c|}{$3.00 \times 10^{-3}$}  & & $3.29 \times 10^{-3}$& $1.49 \times 10^{-4}$ \\
\\
\multirow{3}{*}{\rotatebox[origin=c]{90}{Aug 6}} &\multicolumn{1}{c|}{$\epsilon_r$} & $2.32 \times 10^{-3}$ & $2.78 \times 10^{-3}$ & $2.21 \times 10^{-3}$ & $2.33 \times 10^{-3}$ & \multicolumn{1}{c|}{$2.53 \times 10^{-3}$} \\ 
&\multicolumn{1}{c|}{$\beta$} &$9.59 \times 10^{-4}$ & $1.42 \times 10^{-3}$ & $1.77 \times 10^{-3}$ & $1.36 \times 10^{-3}$ &  \multicolumn{1}{c|}{$1.36 \times 10^{-3}$} \\
&\multicolumn{1}{c|}{$\epsilon$}& $2.51 \times 10^{-3}$ & $3.12 \times 10^{-3}$ & $2.83 \times 10^{-3}$ & $2.70 \times 10^{-3}$ &  \multicolumn{1}{c|}{$2.87 \times 10^{-3}$} & & $2.81 \times 10^{-3}$& $2.01 \times 10^{-4}$ \\
\\
\multirow{3}{*}{\rotatebox[origin=c]{90}{Aug 7}} &\multicolumn{1}{c|}{$\epsilon_r$} &$1.50 \times 10^{-3}$ & $2.37 \times 10^{-3}$ & $2.84 \times 10^{-8}$ & $3.32 \times 10^{-3}$ &  \multicolumn{1}{c|}{$1.74 \times 10^{-8}$} \\ 
&\multicolumn{1}{c|}{$\beta$}& $1.06 \times 10^{-3}$ & $2.38 \times 10^{-3}$ & $4.10 \times 10^{-3}$ & $1.53 \times 10^{-3}$ &  \multicolumn{1}{c|}{$3.11 \times 10^{-3}$} \\
&\multicolumn{1}{c|}{$\epsilon$} &$1.84 \times 10^{-3}$ & $3.36 \times 10^{-3}$ & $4.10 \times 10^{-3}$ & $3.66 \times 10^{-3}$ &  \multicolumn{1}{c|}{$3.11 \times 10^{-3}$}& & $3.21 \times 10^{-3}$& $7.62 \times 10^{-4}$ \\
\\
\cline{1-7} \cline{9-10}
\multirow{4}{*}{\rotatebox[origin=c]{90}{\parbox[r]{1.5 cm}{{\textbf{Time  averaged}}}}} &&&&&&&& \multicolumn{2}{|c|}{\textbf{Total average}} \\
& \multicolumn{1}{c|}{$\boldsymbol{\langle \epsilon \rangle_t}$}
&$2.71 \times 10^{-3}$ & $3.32 \times 10^{-3}$ & $3.37 \times 10^{-3}$ & $3.27 \times 10^{-3}$ & \multicolumn{1}{c|}{$3.11 \times 10^{-3}$} & & \multicolumn{2}{|c|}{$\boldsymbol{\langle \epsilon \rangle}= 3.16\times 10^{-3}$}
\\
& \multicolumn{1}{c|}{$\boldsymbol {\sigma_t}$}& $5.88 \times 10^{-4}$ & $1.15 \times 10^{-4}$  & $4.61 \times 10^{-4}$ & $3.54 \times 10^{-4}$ & \multicolumn{1}{c|}{$2.21 \times 10^{-4}$} & &\multicolumn{2}{|c|}{$\boldsymbol{\langle \sigma_t \rangle_q}= 3.48\times 10^{-4}$}  
\\
&&&&&&&&\multicolumn{2}{|c|}{$\boldsymbol{\langle \sigma_q \rangle_t}= 3.05\times 10^{-4}$}
\\
\cline{1-7} \cline{9-10}
\end{tabular}
\caption{Same as in~\cref{table_1} but for the case of parallel gate implementation of the dephasing-robust pulse.}
\end{table*}

\begin{table*}\centering
\begin{tabular}{@{}l@{\hspace{-1ex}}rcccccc@{}cc}
\cline{1-7} \cline{9-10}
 &&\multicolumn{5}{c}{\textbf{Qubit}} & & \multicolumn{2}{c}{\textbf{System-averaged data}}  \\
 \cline{3-7}
 & & 0 & 1 &2 &3 &4 &&$\boldsymbol{\langle \epsilon \rangle_q}$&$\boldsymbol {\sigma_q}$\\
\cline{1-7} \cline{9-10} \\
\multirow{3}{*}{\rotatebox[origin=c]{90}{Aug 4}} & \multicolumn{1}{c|}{$\epsilon_r$} &
$3.79 \times 10^{-8}$ & $7.85 \times 10^{-8}$ & $6.93 \times 10^{-8}$ & $6.53 \times 10^{-3}$ & 
\multicolumn{1}{c|}{$6.07 \times 10^{-3}$} & & &  \\
 &\multicolumn{1}{c|}{$\beta$} & $3.10 \times 10^{-3}$ & $6.03 \times 10^{-3}$ & $5.77 \times 10^{-3}$ & $3.22 \times 10^{-3}$ & \multicolumn{1}{c|}{$3.16 \times 10^{-3}$} & & & \\
&\multicolumn{1}{c|}{$\epsilon$} &$3.10 \times 10^{-3}$ & $6.03 \times 10^{-3}$ & $5.77 \times 10^{-3}$ & $7.28 \times 10^{-3}$ & \multicolumn{1}{c|}{$6.84 \times 10^{-3}$} & & $5.81 \times 10^{-3}$& $1.46 \times 10^{-3}$   \\
\\
\multirow{3}{*}{\rotatebox[origin=c]{90}{Aug 5}} &\multicolumn{1}{c|}{$\epsilon_r$} & $5.43 \times 10^{-9}$ & $1.04 \times 10^{-7}$ & $4.16 \times 10^{-8}$ & $6.98 \times 10^{-3}$ & \multicolumn{1}{c|}{$5.96 \times 10^{-3}$} &&& \\ 
&\multicolumn{1}{c|}{$\beta$} & $2.79 \times 10^{-3}$ & $5.82 \times 10^{-3}$ & $5.22 \times 10^{-3}$ & $3.64 \times 10^{-3}$ & \multicolumn{1}{c|}{$4.56 \times 10^{-3}$} &&&\\
&\multicolumn{1}{c|}{$\epsilon$} & $2.79 \times 10^{-3}$ & $5.82 \times 10^{-3}$ & $5.22 \times 10^{-3}$ & $7.87 \times 10^{-3}$ & \multicolumn{1}{c|}{$7.50 \times 10^{-3}$}  & & $5.84 \times 10^{-3}$& $1.82 \times 10^{-3}$ \\
\\
\multirow{3}{*}{\rotatebox[origin=c]{90}{Aug 6}} &\multicolumn{1}{c|}{$\epsilon_r$} & $4.66 \times 10^{-8}$ & $1.13 \times 10^{-7}$ & $2.56 \times 10^{-8}$ & $6.45 \times 10^{-3}$ & \multicolumn{1}{c|}{$6.50 \times 10^{-3}$} \\ 
&\multicolumn{1}{c|}{$\beta$} &$2.76 \times 10^{-3}$ & $5.37 \times 10^{-3}$ & $5.16 \times 10^{-3}$ & $3.13 \times 10^{-3}$ & \multicolumn{1}{c|}{$3.21 \times 10^{-3}$} \\
&\multicolumn{1}{c|}{$\epsilon$}& $2.76 \times 10^{-3}$ & $5.37 \times 10^{-3}$ & $5.16 \times 10^{-3}$ & $7.17 \times 10^{-3}$ & \multicolumn{1}{c|}{$7.25 \times 10^{-3}$} & & $5.54 \times 10^{-3}$& $1.64 \times 10^{-3}$ \\
\\
\multirow{3}{*}{\rotatebox[origin=c]{90}{Aug 7}} &\multicolumn{1}{c|}{$\epsilon_r$} &$1.96 \times 10^{-7}$ & $9.77 \times 10^{-9}$ & $3.37 \times 10^{-8}$ & $6.87 \times 10^{-3}$ & \multicolumn{1}{c|}{$6.05 \times 10^{-3}$} \\ 
&\multicolumn{1}{c|}{$\beta$}& $2.91 \times 10^{-3}$ & $6.43 \times 10^{-3}$ & $7.81 \times 10^{-3}$ & $3.43 \times 10^{-3}$ & \multicolumn{1}{c|}{$5.68 \times 10^{-3}$} \\
&\multicolumn{1}{c|}{$\epsilon$} &$2.91 \times 10^{-3}$ & $6.43 \times 10^{-3}$ & $7.81 \times 10^{-3}$ & $7.68 \times 10^{-3}$ & \multicolumn{1}{c|}{$8.30 \times 10^{-3}$}& & $6.63 \times 10^{-3}$& $1.96 \times 10^{-3}$ \\
\\
\cline{1-7} \cline{9-10}
\multirow{4}{*}{\rotatebox[origin=c]{90}{\parbox[r]{1.5 cm}{{\textbf{Time  averaged}}}}} &&&&&&&& \multicolumn{2}{|c|}{\textbf{Total average}} \\
& \multicolumn{1}{c|}{$\boldsymbol{\langle \epsilon \rangle_t}$}
&$2.89 \times 10^{-3}$ & $5.91 \times 10^{-3}$ & $5.99 \times 10^{-3}$ & $7.50 \times 10^{-3}$ & \multicolumn{1}{c|}{$7.47 \times 10^{-3}$} & & \multicolumn{2}{|c|}{$\boldsymbol{\langle \epsilon \rangle}= 5.95\times 10^{-3}$}
\\
& \multicolumn{1}{c|}{$\boldsymbol {\sigma_t}$}& $1.32 \times 10^{-4}$ & $3.84 \times 10^{-4}$  & $1.08 \times 10^{-3}$ & $2.85 \times 10^{-4}$ & \multicolumn{1}{c|}{$5.31 \times 10^{-4}$} & &\multicolumn{2}{|c|}{$\boldsymbol{\langle \sigma_t \rangle_q}= 4.82\times 10^{-4}$}  
\\
&&&&&&&&\multicolumn{2}{|c|}{$\boldsymbol{\langle \sigma_q \rangle_t}= 1.72\times 10^{-3}$}
\\
\cline{1-7} \cline{9-10}
\end{tabular}
\caption{Same as in~\cref{table_1} but for the case of parallel gate implementation of the dual robust pulse.}
\end{table*}

\end{widetext}

\end{document}